\input pipi.sty
\input epsf.sty
\magnification1020
\raggedbottom

\nopagenumbers
\rightline\timestamp
\rightline{FTUAM 03-1}
\rightline{hep-ph/0304067}
\rightline{June, 30, 2003}
\bigskip
\hrule height .3mm
\vskip.6cm
\centerline{{\bigfib On the precision of chiral-dispersive calculations
 of $\pi\pi$ scattering}\footnote{}{\petit$^{(*)}$ On 
leave from Departamento de F\'{\i}sica Te\'orica,~II
 (M\'etodos Matem\'aticos),
Facultad de Ciencias F\'{\i}sicas,
Universidad Complutense de Madrid.}}
\medskip
\centerrule{.7cm}
\vskip1cm
\setbox8=\vbox{\hsize65mm {\noindent\fib J. R. Pel\'aez} 
\vskip .1cm
\noindent{\addressfont INFN, Sezione di Firenze and\hb
 Dipartimento di Fisica,\hb
 Universit\`a degli Studi,\hb
 I-50019, Sesto Fiorentino. (FI), Italy.$^{(*)}$}}
\centerline{\box8}
\smallskip
\setbox7=\vbox{\hsize65mm \fib and} 
\centerline{\box7}
\smallskip
\setbox9=\vbox{\hsize65mm {\noindent\fib F. J. 
Yndur\'ain} 
\vskip .1cm
\noindent{\addressfont Departamento de F\'{\i}sica Te\'orica, C-XI\hb
 Universidad Aut\'onoma de Madrid,\hb
 Canto Blanco,\hb
E-28049, Madrid, Spain.}\hb}
\smallskip
\centerline{\box9}
\bigskip
\setbox0=\vbox{\abstracttype{Abstract}We calculate the combination $2a_0^{(0)}-5a_0^{(2)}$ 
(the Olsson sum rule) and 
the scattering lengths and effective ranges $a_1$, $a_2^{(I)}$ and $b_1$, $b_2^{(I)}$
 dispersively (with the Froissart--Gribov representation) using, at low energy, 
the phase shifts for $\pi\pi$ scattering obtained by Colangelo, Gasser and Leutwyler 
(CGL) from 
the Roy equations and 
chiral perturbation theory, plus experiment and Regge behaviour at high energy, 
or directly, using the CGL parameters for $a$s and $b$s. 
We find  mismatch, both among the CGL phases themselves and with the 
results obtained from the pion form factor. 
This reaches the level of several ($2$ to $5$) standard deviations,
and is essentially independent of the  
details of the intermediate energy region ($0.82\leq E\leq 1.42$ GeV) and, in some cases,    
 of the high energy behaviour assumed. 
We discuss possible reasons for this  mismatch, 
in particular in connection with an alternate set of phase shifts.
}
\centerline{\box0}
\brochureendcover{Typeset with \physmatex}
\brochureb{\smallsc j. r. pel\'aez and f. j.  yndur\'ain}{\smallsc on the precision
 of chiral-dispersive calculations
 of $\pi\pi$ scattering}{1}

\brochuresection{1. Introduction}

\noindent
In two remarkable recent papers, Ananthanarayan, Colangelo, Gasser and Leutwyler\ref{1} 
and Colangelo, Gasser and Leutwyler\ref{2} (to be referred to as, respectively, 
ACGL and CGL) have used 
experimental information, analyticity and unitarity (in the form of  
the Roy equations\ref{3}) and, in CGL, chiral calculations to two loops, 
to construct what is presented as a very precise $\pi\pi$ 
scattering amplitude at low energy, 
$E\equiv s^{1/2}\leq0.8\,\gev$.

There is little doubt that the small errors claimed by CGL, at the level 
of very few percent, follow from the Roy-chiral analysis,
 plus chiral perturbation theory(with the assumption
of  negligible higher order corrections),
given the input scattering amplitude at high energy, 
say, for $E\gsim1.42\,\gev$. 
What is however not so clear is that the input selected by ACGL 
is unique, not even that it is the more physically acceptable one. 
The question then remains, what is the effect of changing this high energy 
input in the low energy $\pi\pi$ amplitude.

In the present paper we address ourselves to the matter of the 
consistency of the CGL $S$ matrix. 
To be precise, we evaluate the following quantities: 
the combination of S0, S2 scattering 
lengths $2a_0^{(0)}-5a_0^{(2)}$ ({\sl Olsson sum rule}); 
the scattering length $a_1$ and effective range\fnote{Actually, 
$b_1$ is not the effective range, though it is related to it. 
We use the definitions of ACGL and CGL for 
the $a$ and $b$, except that we 
take the dimensions of $a_l$ to be $M_{\pi}^{-(2l+1)}$. 
Here $M_\pi$ is the charged pion mass, $M_\pi\simeq139.57\,\mev$.}
 $b_1$ in the P wave; and the scattering lengths and 
effective ranges $a_2^{(I)},\;b_2^{(I)}\;I=0,\,2$ for the D0, D2 
waves. For the $a_1$, $b_1$,  $a_2^{(I)}$, $b_2^{(I)}$
 we use the Froissart--Gribov representation.\fnote{The 
method of the Froissart--Gribov representation to calculate 
scattering lengths and effective rages 
was introduced in refs.~4,\/5. 
It is also discussed in some detail in ref.~6.}
This presents two advantages. 
First of all, it was {\sl not} verified 
in ACGL or CGL; 
therefore, it provides a novel test of the CGL phase shifts. 
Secondly, for $a_1,\,b_1$ and, to a lesser extent, for the $a_2^{(I)}$, 
the Froissart--Gribov representation is 
sensitive to the high energy scattering amplitude,
 precisely 
one of the features
 we want to probe.
We then compare what we find with the values for $a$s and $b$s given by CGL 
themselves.
For $a_1,\,b_1$ we also compare the CGL evaluations with 
the results of a direct fit of the P wave to the  
pion form factor, this last a fully independent test.

The result of our calculations (\sect~4 here) is that the solution of CGL is not consistent with 
the results from the fit to the pion form factor or with 
itself (if assuming a reasonable high energy Regge behaviour)  and 
the mismatch occurs essentially independently of 
the details of the intermediate energy ($0.82\leq s^{1/2}\leq1.42$ \gev)
 phase shifts we use, provided they fit experiment (\subsect~4.4), and,
in some cases,
also of
 assumptions on the high energy ($s^{1/2}\geq1.42$ \gev) behaviour. 
For some of the quantities discussed above
the disagreement reaches several (up to 5) standard deviations.
For the $a_2^{(I)}$ the more striking discrepancy 
occurs for the combination $a_{0+}=\tfrac{2}{3}[a_2^{(0)}-a_2^{(2)}]$. 
This is because it corresponds to a combination of only 
isospin 1, 2 in the $s$, $u$ channels so the Froissat--Gribov
integral is very accurate since te S0 wave, large and the one less well known, does not 
contribute. 
The chiral perturbation theory calculation for this quantity has also small errors 
since (to one loop) it only depends on one chiral lagrangian constant, $\bar{l}_2$, 
see below.

The mismatch is much less severe (below the $2\,\sigma$ level)
 for the ACGL results, the main reason being
that their 
errors are at least three times as large as the CGL ones. We discuss in Sect.~5 
the reasons for the CGL mismatch, which may be due to the use by CGL of 
an irrealistic high energy part of the scattering amplitude, 
which distorts their low energy ($s^{1/2}<0.82$ GeV) 
phase shifts beyond the 
very small errors implied by their assumption of negligible
 higher chiral perturbative corrections.

Apart from these two sections, we present in \sect~2 the Roy equations, 
in \sect~3 the scattering amplitude we will use (including in particular a detailed 
discussion of the high energy pieces) and finish the
 paper with  summary and conclusions in \sect~6.

\vfill\eject
\brochuresection{2. The Roy equations}
\vskip-0.5truecm
\brochuresubsection{2.1. Dispersion relations}

\noindent The analyticity properties of the $\pi\pi$ scattering amplitude, $F(s,t)$, 
 imply that we can write a Cauchy
representation for it, fixing $t$ and allowing $s$ to be complex. 
 For $s$ physical this reads
$$\real F(s,t)=D(s,t)=\dfrac{1}{\pi}\pepe\int_{4M_{\pi}^2}^\infty\dd s'\,
\dfrac{A_s(s',t)}{s'-s}+\dfrac{1}{\pi}\int_{4M_{\pi}^2}^\infty\dd s'\,\dfrac{A_u(s',t)}{s'-u},
\quad A(s,t)=\imag F(s,t).
\equn{(2.1)}$$
($\pepe$ denotes Cauchy's principal part of the integral).

Actually, and because, in some cases, the $A(s,t)$ grow 
linearly with $s$, (2.1) is divergent. 
This is repaired by {\sl subtractions}; that is to say,
by writing the Cauchy representation not for $F$ itself,
but for 
$F(s,t)/(s-s_1)$ where  $s_1$ is a convenient subtraction point, usually taken
 to coincide with a  threshold. 
This introduces a function of $t$ in the equations (the value of $F(s,t)$ at $s=s_1$); 
we leave it to the reader to rewrite our equations with the appropriate subtraction incorporated

Let us  separate out 
the high energy contribution, $s\geq s_h$ (we will fix $s_h$ later) to (2.1). 
We then have
$$D(s,t)=\dfrac{1}{\pi}\pepe\int_{4M_{\pi}^2}^{s_h}\dd s'\,
\dfrac{A_s(s',t)}{s'-s}+\dfrac{1}{\pi}\int_{4M_{\pi}^2}^{s_h}\dd s'\,\dfrac{A_u(s',t)}{s'-u}+
V(s,t;s_h)
\equn{(2.2a)}$$
and
$$V(s,t;s_h)=\dfrac{1}{\pi}\int_{s_h}^\infty\dd s'\,
\dfrac{A_s(s',t)}{s'-s}+\int^{\infty}_{s_h}\dd s'\,\dfrac{A_u(s',t)}{s'-u};
\equn{(2.2b)}$$
we are assuming $s<s_h$. 
Both $D$ and the $A$ may be written in terms of the {\sl same} set of phase shifts
by expanding them,\fnote{We are actually simplifying in that  
(2.2) should take into account the different isospin structure of $s$ and $u$ channels, 
which the reader may find in e.g. the text of Martin, Morgan and Shaw.\ref{7}}
for fixed $s$ channel isospin $I$, as
$$A^{(I)}(s,t)=2\dfrac{2s^{1/2}}{\pi k}\sum_{l=0}^\infty(2l+1)P_l(\cos\theta)
\dfrac{1}{\cot^2\delta_l^{(I)}(s)+1},
\equn{(2.3a)}$$ 
$$D^{(I)}(s,t)=
2\dfrac{2s^{1/2}}{\pi k}\sum_{l=0}^\infty(2l+1)P_l(\cos\theta)
\dfrac{\cot\delta_l^{(I)}(s)}{\cot^2\delta_l^{(I)}(s)+1}.
\equn{(2.3b)}$$
One of the factors 2 above occurs because of the identity of pions; 
we work in the limit of exact isospin invariance.

These equations provide {\sl constraints} for the phase shifts 
provided one knows (or has a reliable model) for the 
high energy term, $V(s,t;s_h)$. 
They enforce analyticity and $s\leftrightarrow u$ crossing symmetry.

\brochuresubsection{2.2. The Roy equations}

\noindent
Eqs.~(2.1) to (2,3) look rather cumbersome. Roy\ref{3} remarked that they 
appear simpler if we project them  into partial waves, 
integrating over physical ($t\leq0$) values of the cosine of the scattering angle: 
one finds the {\sl Roy equations}
$$\dfrac{\cot\delta_l^{(I)}(s)}{\cot^2\delta_l^{(I)}(s)+1}=
\sum_{l'=0}^\infty\int_{4M_{\pi}^2}^{s_h}\dd s'\,
K_{ll'}(s,s')\dfrac{1}{\cot^2\delta_{l'}^{(I)}(s')+1}+V_l(s;s_h).
\equn{(2.4)}$$
Here the kernels $K_{ll'}$ are known and the $V_l$ are the  
projections of $V$.

Eq.~(2.4) is valid in 
the simplified case we are considering here, i.e., without subtractions. 
If we had subtractions, the fixed $t$ dispersion 
relations would acquire an extra term, a function $g(t)$ (the value of 
$F(s_1,t)$ at the subtraction point). 
This may be eliminated, using crossing symmetry, 
in favour of the S wave 
scattering lengths. 
\equn{(2.4)} would be modified accordingly.

Let us rewrite the Roy equations in the form
$$\xi=\phiv(\xi,V)
\equn{(2.5)}$$
where $\xi=\{\imag f_l\}_{l=0}^{\infty}$ 
stands for the set of imaginary parts of the partial waves, 
for $s\leq s_h$, and $\phiv$ 
is the functional that follows 
from (2.4). 
We can define a mapping,
$$\xi'\equiv\phiv(\xi,V)
\equn{(2.6)}$$ 
and then the solution of the Roy equations is a fixed point of $\phiv$.
 
The relations (2.5) are highly nonlinear integral and matrix equations. 
Solutions are known to exist in some favorable cases; 
in fact, Atkinson\ref{8} proved, even before the 
advent of Roy's equations,  that, for any arbitrary $V(s,t;s_h)$ 
such that it is sufficiently smooth and decreasing at infinity, 
one can obtain, by iterating (2.6), a solution not 
only of the Roy equations, but of the full Mandelstam representation, and 
compatible with inelastic unitarity for all $s$ as well.
Therefore, the solutions to the Roy equations  are ambiguous in an 
unknown function, and the matter of what is an acceptable $V$ 
becomes crucial.
 This is particularly so because fulfillment of the Roy
 equations does {\sl not} guarantee
full analyticity and crossing; and it may happen that a
 given solution of the Roy equations is
incompatible with other sum rules (as is the case for the CGL solution).

\brochuresection{3. The scattering amplitude}

\noindent
At low energy, say $s^{1/2}\leq0.82\, \gev$, the inelasticity in $\pi\pi$ scattering 
is known experimentally to be negligible; 
it is for these energies that the Roy equations (2.5) are to be solved. 
To do so we need as input the function $V$  or, equivalently, 
the imaginary part of the scattering amplitude for 
 energies $s^{1/2}\geq 0.82\,\gev$. 
In fact, for the Roy equations we need $\imag F(s,t)$ for $s$ physical 
 and $t$ physical, $t\leq0$. 
However, for other applications, 
we will require $\imag F(s,t)$ up to the edge of the Martin--Lehmann 
ellipse,\fnote{For analyticity properties of $\pi\pi$ scattering see for example 
ref.~7.} $t\leq 4M^2_\pi$; our discussion will also cover this case.
We now proceed with a discussion of the different waves and energy regions.

\brochuresubsection{3.1. The S and P waves for $E$ below $0.82$ \gev}

\noindent
Because we want to test the solution of CGL for the $\pi\pi$ $S$ matrix, we consider now 
the solution to the Roy 
equations,  incorporating chiral perturbation theory to two loops, given there. 
The low energy S0, S2, P waves are written by these authors as
$$\tan\delta_l^{(I)}(s)=k^{2l}\sqrt{1-4M^2_\pi/s}
\left\{A_l^I+B_l^Ik^2/M^2_\pi+C_l^Ik^4/M^4_\pi+D_l^Ik^6/M^6_\pi\right\}\dfrac{4M^2_\pi-s_{lI}}{s-s_{lI}},
\equn{(3.1a)}$$
$k=\sqrt{s/4-M^2_\pi}$, and the values of the parameters, as given by CGL, \equn{(17.2)},  are
$$\matrix{
A_0^0=0.220,&B_0^0=0.268,&C_0^0=-0.0139,&D_0^0=-0.139/10^2,&s_{00}=36.77\,M^2_\pi\cr
A_0^2=-0.444/10,&B_0^2=-0.857/10,&C_0^2=-0.221/10^2,&D_0^2=-0.129/10^3,&s_{02}=-21.62\,M^2_\pi\cr
A_1=0.379/10,&B_1=0.140/10^4,&C_1=-0.673/10^4,&D_1=0.163/10^7,&s_1=30.72\,M^2_\pi.\cr
}
\equn{(3.1b)}$$
These are the values of the phase shifts that we will use up to the energy $E=0.82\,\gev$. 

To test dispersion relations, either in the form of 
the Olsson relation or the Froissart--Gribov representation, 
we need also the values of the S, P waves at intermediate energies 
($0.82\leq E\leq 1.42$) and the values of the 
D, F waves below 1.42 \gev, that we take from experiment; 
higher waves are presumably negligible. 
Moreover, we require $\imag F(s,t)$ for $s^{1/2}\geq1.42\,\gev$. 
This last we will obtain from Regge theory in \subsect~3.4; 
we now turn to the intermediate energy regions.

Before doing so, however, we want to emphasize that, in the present paper,
 we do {\sl not} deal with 
the matter of the consistency of the fits for the S, P waves between 0.82 and 1.42 \gev\ 
that we will give in next subsection, or of those for the D, F waves. 
These fits [Eqs.~(3.2) to (3.11) below] are merely a convenient way to 
summarize the {\sl experimental} data; our results 
would change very little if we had instead used a spline interpolation for  
the experimental phase shifts. 
We will discuss this further in \subsect~4.4.1, 
where we will show that the discrepancy will remain essentially unchanged provided 
we demand a resemblance to the data (allowing for a large uncertainty)
 in this intermediate region.

\brochuresubsection{3.2. The S, P waves between 0.82 and 1.42 \gev}

\noindent
For the S0 wave in the region between 0.82 \gev\ and $\bar{K}K$ threshold we use the parametrization, 
obtained by fitting experimental data\fnote{The (slight) differences with some of the 
parameters in ref.~6 occur because now we are using $M_\pi=m_{\pi^+}=139.57\,\mev$
 instead of the average pion mass, 
138 \mev, and we are also essentially eliminating from 
the fit the data for energies above 0.96 \mev. 
The change in the \chidof\ of the $I=0$ S wave corrects an error there. 
We send to this reference for details on the fitting procedure.} (as in ref.~6),
$$\eqalign{
\cot\delta_0^{(0)}(s)=&\,\dfrac{s^{1/2}}{2k}\,\dfrac{M_{\pi}^2}{s-\tfrac{1}{2}M_{\pi}^2}\,
\dfrac{M^2_\sigma-s}{M^2_\sigma}\,
\left\{B_0+B_1\dfrac{\sqrt{s}-\sqrt{s_0-s}}{\sqrt{s}+\sqrt{s_0-s}}+
B_2\left[\dfrac{\sqrt{s}-\sqrt{s_0-s}}{\sqrt{s}+\sqrt{s_0-s}}\right]^2\right\};\cr
s_0^{1/2}=2M_K;&\quad\chi^2/{\rm d.o.f.}=11.1/(19-4).\cr
\quad M_\sigma=806\pm21,&\,\quad B_0=21.91\pm0.62,\quad B_1=20.29\pm1.55, \quad B_2=22.53\pm3.48;\cr
a_0^{(0)}=&\,(0.226\pm0.015)\;M_{\pi}^{-1}.\cr
\cr
}
\equn{(3.2)}$$

\topinsert{
\setbox0=\vbox{\hsize13.truecm{\epsfxsize 11.truecm\epsfbox{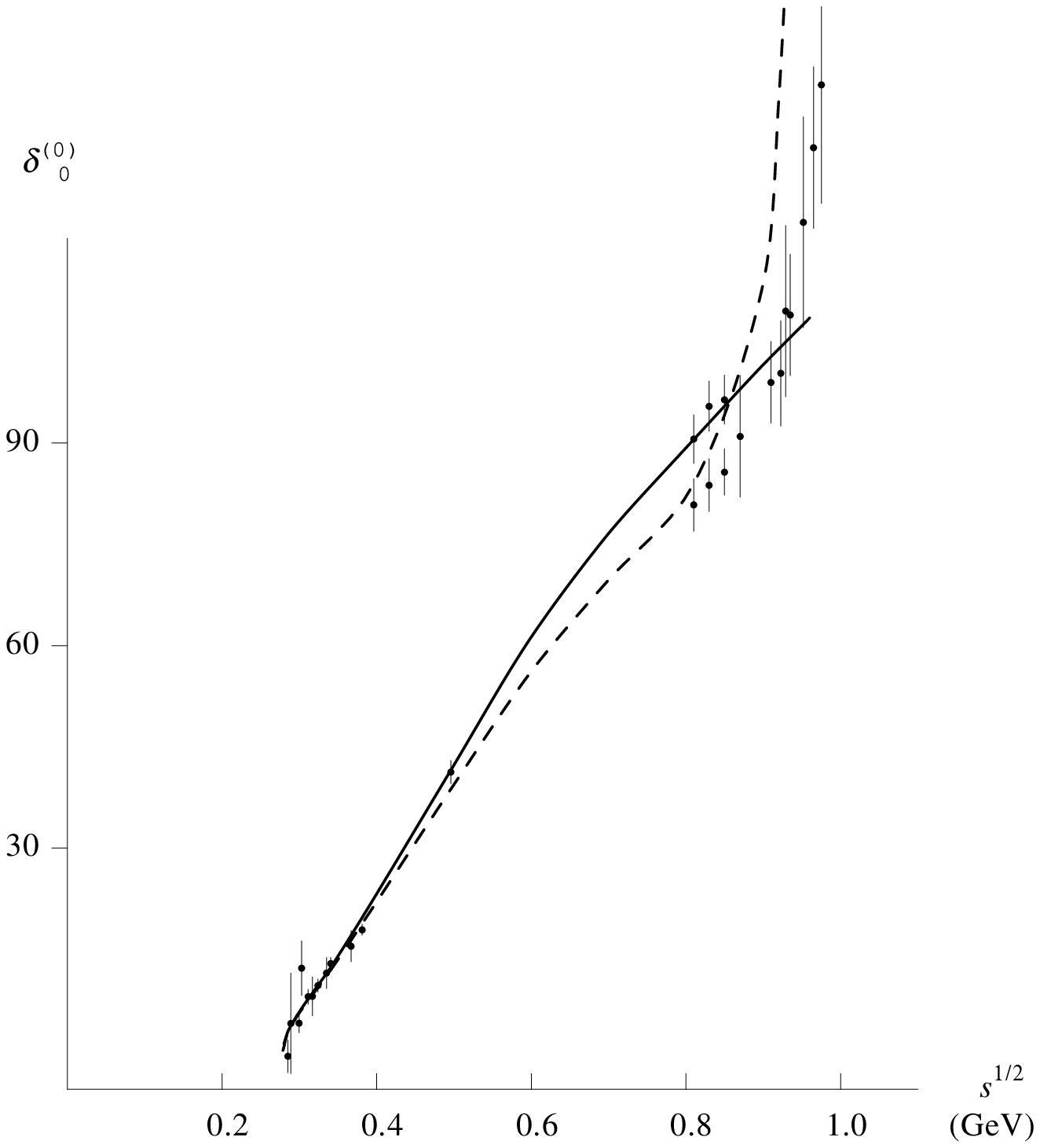}}} 
\setbox6=\vbox{\hsize 11.4truecm\captiontype\figurasc{Figure 3.1. }{
The  
$I=0$, $S$-wave phase shifts corresponding to (3.2) (continuous line) and 
 Colangelo, Gasser and Leutwyler, ref.~2 (dashed line).}\hb} 
\centerline{\tightboxit{\box0}}
\bigskip
\centerline{\box6}
}\endinsert
The  solution depends on the value of $\delta_0^{(0)}(M_K^2)$ we impose in the fit. 
In (3.2) we took that following from the more recent measurements of 
$K_{2\pi}$ decay,\ref{9} $\delta_0^{(0)}(M_K^2)=41.5\pm3\degrees$; 
another possibility  
is to average this with the older determination,\ref{9} 
thus imposing the value $\delta_0^{(0)}(M_K^2)=43.3\pm2.3\degrees$; 
this we will discuss in \subsect~5.2.

Solution (3.2) is, up to $s^{1/2}\sim0.84\,\gev$, similar to the CGL one, (3.1); 
see \fig~3.1.  We will use the CGL solution up to $0.82\,\gev$, 
slightly above 
 their nominal maximum range, $s^{1/2}=0.80\,\gev$, 
and (3.2) between 0.82 and 0.96 \gev.

For the S2 wave between 0.82 \gev\ and 1.42 \gev\ we use the phase shift  
obtained by fitting experimental data and including the requirement 
$a_0^{(2)}=0.044\pm0.003\,M^{-1}_\pi$ (this last follows from the analysis of CGL):
\smallskip
$$\eqalign{
\cot\delta_0^{(2)}(s)=&\,\dfrac{s^{1/2}}{2k}\,\dfrac{M_{\pi}^2}{s-2z_2^2}\,
\left\{B_0+B_1\dfrac{\sqrt{s}-\sqrt{s_0-s}}{\sqrt{s}+\sqrt{s_0-s}}\right\};\cr
s_0^{1/2}=1.45\;\gev;&\quad\chi^2/{\rm d.o.f.}=16.1/(18-2).\cr
 B_0=&\,-115\pm4,\quad B_1=-106\pm3,\quad z_2=139.57\;\mev\;\hbox{[fixed]}.\cr
}
\equn{(3.3)}$$
This actually corresponds to $a_0^{(2)}=-0.0457\pm0.0074$. 
One can allow  
$z_2$ to vary by 8 \mev, and still be within $1\,\sigma$ of the minimum,
 but we will not do so here.

Then we have the P wave between 0.82 \gev\ and 1.0 \gev. 
Here we fit the pion form factor, including $e^+e^-$ and $\tau$ decay data. 
There are now two possibilities: the first one is
\smallskip
$$\eqalign{
\cot\delta_1(s)=&\,\dfrac{s^{1/2}}{2k^3}\,(M^2_\rho-s)\,
\left\{B_0+B_1\dfrac{\sqrt{s}-\sqrt{s_0-s}}{\sqrt{s}+\sqrt{s_0-s}}\right\};\cr
s_0^{1/2}=1.05\;\gev;&\quad\chi^2/{\rm d.o.f.}=1.3.\cr
\quad M_\rho=772.3\pm0.5\;\mev,&\,\quad B_0=1.060\pm0.005,\quad B_1=0.24\pm0.06.\cr
[0.82\,\gev\leq s^{1/2}\leq&\, 1.0\;\gev].\cr
}
\equn{(3.4)}$$
In particular, for the low energy parameters, this gives
$$a_1=(40.6\pm1.3)\times10^{-3}\;M_{\pi}^{-3},\quad b_1=(4.18\pm0.43)\times10^{-3}\;M_{\pi}^{-5}.$$
This result is  obtained from the fit to the pion form factor, with only statistical 
experimental errors taken into account, performed in 
ref.~10. If we also take 
systematic normalization errors into account, (3.4) is replaced by
\smallskip
$$\eqalign{
\cot\delta_1(s)=&\,\dfrac{s^{1/2}}{2k^3}\,(M^2_\rho-s)\,
\left\{B_0+B_1\dfrac{\sqrt{s}-\sqrt{s_0-s}}{\sqrt{s}+\sqrt{s_0-s}}\right\};\cr
s_0^{1/2}=1.05\;\gev;&\quad\chi^2/{\rm d.o.f.}=1.1.\cr
\quad M_\rho=773.5\pm0.85\;\mev,&\,\quad B_0=1.071\pm0.007,\quad B_1=0.18\pm0.05.\cr
[0.82\,\gev\leq s^{1/2}\leq&\, 1.0\;\gev]\cr
}
\equn{(3.5)}$$
and now
$$a_1=(38.6\pm1.2)\times10^{-3}\;M_{\pi}^{-3},\quad b_1=(4.47\pm0.29)\times10^{-3}\;M_{\pi}^{-5}.$$
We will consider both possibilities, but the calculations of 
dispersive and Froissart--Gribov integrals will be made with (3.5), 
for definiteness. 
If using (3.4) the differences would be minute.\fnote{
The fact  that the  errors for $a_1$, 
$b_1$ are smaller when using (3.5) than when using (3.4), which at first sight appears 
 counterintuitive, can be understood as follows. 
The errors in the parameters $B_0$, $M_\rho$ in (3.5) are larger than those in 
(3.4) --as suggested by intuition. The error in $B_1$, however, is smaller, and 
it is his quantity that influences most  
$b_1$ (the error in $a_1$ stays 
essentially constant).  
Including systematic errors makes the determinations of the pion form factor in the 
timelike and spacelike regions more compatible one with another, 
and this allows a more precise determination of low 
energy parameters.}

\topinsert{
\setbox0=\vbox{\hsize10.truecm{\epsfxsize 8.2truecm\epsfbox{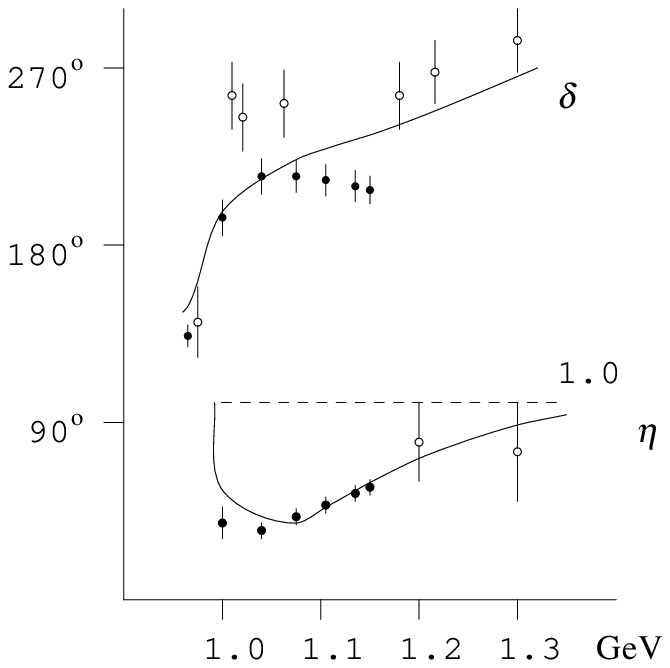}}} 
\setbox6=\vbox{\hsize 11truecm\captiontype\figurasc{Figure 3.2. }{
Fits to the  
$I=0$, $S$-wave phase shift and inelasticity from 960 to 1350 \mev. 
Also shown are the data points from solution~1 of Protopopescu et al.\ref{12} (black dots) and 
some data  of  Grayer et al.\ref{11} (open circles).
}\hb} 
\centerline{\tightboxit{\box0}}
\bigskip
\centerline{\box6}
\medskip
}\endinsert

We next turn to the S0, P waves in the higher energy regions, but still $E\leq1.42\,\gev$.
For the S0 wave between $\bar{K}K$ threshold, $0.992\,\gev$, 
and $1.42\,\gev$, we use a semiempirical formula that fits reasonably well 
the existing data,\ref{11,12} from $s^{1/2}\geq0.96\,\gev$ to 1.50 \gev:
$$\eqalign{
\cot\delta_0^{(0)}(s)=&\,c_0\dfrac{(s-M^2_\sigma)(M_f^2-s)|k_2|}{M^2_fs^{1/2}k_2^2};\quad
k_2=\dfrac{\sqrt{s-4M_K^2}}{2}\cr
\eta=&\,1-\left(c_1\dfrac{k_2}{s^{1/2}}+c_2\dfrac{k_2^2}{s}\right)
\dfrac{M'^{\/2}-s}{s},\cr
[0.992\leq s^{1/2}\leq1.42\;\gev]\quad c_0=&\,1.36\pm0.05,\;c_1=6.7\pm0.15,\;c_2=-17.6\pm0.7,\cr
M_K=496\,\mev,\quad M_\sigma=&\,0.802\,\gev,\;M_f=1.32\,\gev,\;M'=1.5\,\gev.
}
\equn{(3.7)}$$
Note that, for inelastic scattering, we define our parameters so that, in general,
$$\imag \hat{f}_l^{(I)}(s)=\dfrac{\eta_l^{(I)}}{1+\cot^2\delta_l^{(I)}(s)}+
\dfrac{1-\eta_l^{(I)}}{2};
$$
in the elastic region, $\eta_l^{(I)}(s)=1$. 
The fit to the data following from (3.7) is shown in Fig.~3.2.

Finally, for the P wave between 1 \gev\ and 1.42 \gev, we use an empirical formula, 
obtained  adding a resonance (with mass 1.45 \gev) to a nonresonant background:
$$\eqalign{
\imag \hat{f}_1(s)=&\,\dfrac{1}{1+[\lambda+1.1k_2/s^{1/2}]^2}+
{\rm BR}\dfrac{M^2_{\rho'}\gammav^2}{(s-M^2_{\rho'})^2+M^2_{\rho'}\gammav^2};\cr
[1.0\leq s^{1/2}\leq1.42\;\gev]\quad M_{\rho'}=&\,1.45\;\gev,
\quad \gammav=0.31\;\gev,\quad \lambda=2.6\pm0.2;\quad {\rm BR}=0.25\pm0.05.\cr }
\equn{(3.8)}$$
Note that the effect of the $\rho(1450)$ is very small, as will be clear in our various evaluations 
below.

\brochuresubsection{3.3. The D, F waves below 1.42 \gev}

\noindent
We take these waves as given (from threshold to 1.42 \gev)
 by the fits of ref.~6, with inelasticity added 
for the D0 wave and, for the F wave, including also the tail of the $\rho_3$ resonance. 
Moreover, we have required (for compatibility with the CGL 
analysis) that the corresponding scattering lengths agree within errors 
with those given in CGL; that is to say, we include the CGL values, 
weighted with their errors, in the fits for D2, F (for the D0 wave it
 is not necessary, as there are enough precise experimental data).
For the D0 wave we thus write
$$\eqalign{
\cot\delta_2^{(0)}(s)=&\,\dfrac{s^{1/2}}{2k^5}\,(M_{f_2}-s)M^2_\pi\,
\left\{B_0+B_1\dfrac{\sqrt{s}-\sqrt{s_0-s}}{\sqrt{s}+\sqrt{s_0-s}}\right\};\quad
s_0^{1/2}=1.430\;\gev;\cr
\quad M_{f_2}=1270\;\mev\;,&\,\quad B_0=23.7\pm0.7,\quad B_1=22.9\pm2.7.\cr
\eta=&\,1-2\times0.15\dfrac{2[k/k(M^2_{f_2})]^{10}}{1+[k/k(M^2_{f_2})]^{20}}.
\cr
}
\equn{(3.9)}$$

\topinsert{
\setbox0=\vbox{\hsize13.truecm{\epsfxsize 10.8truecm\epsfbox{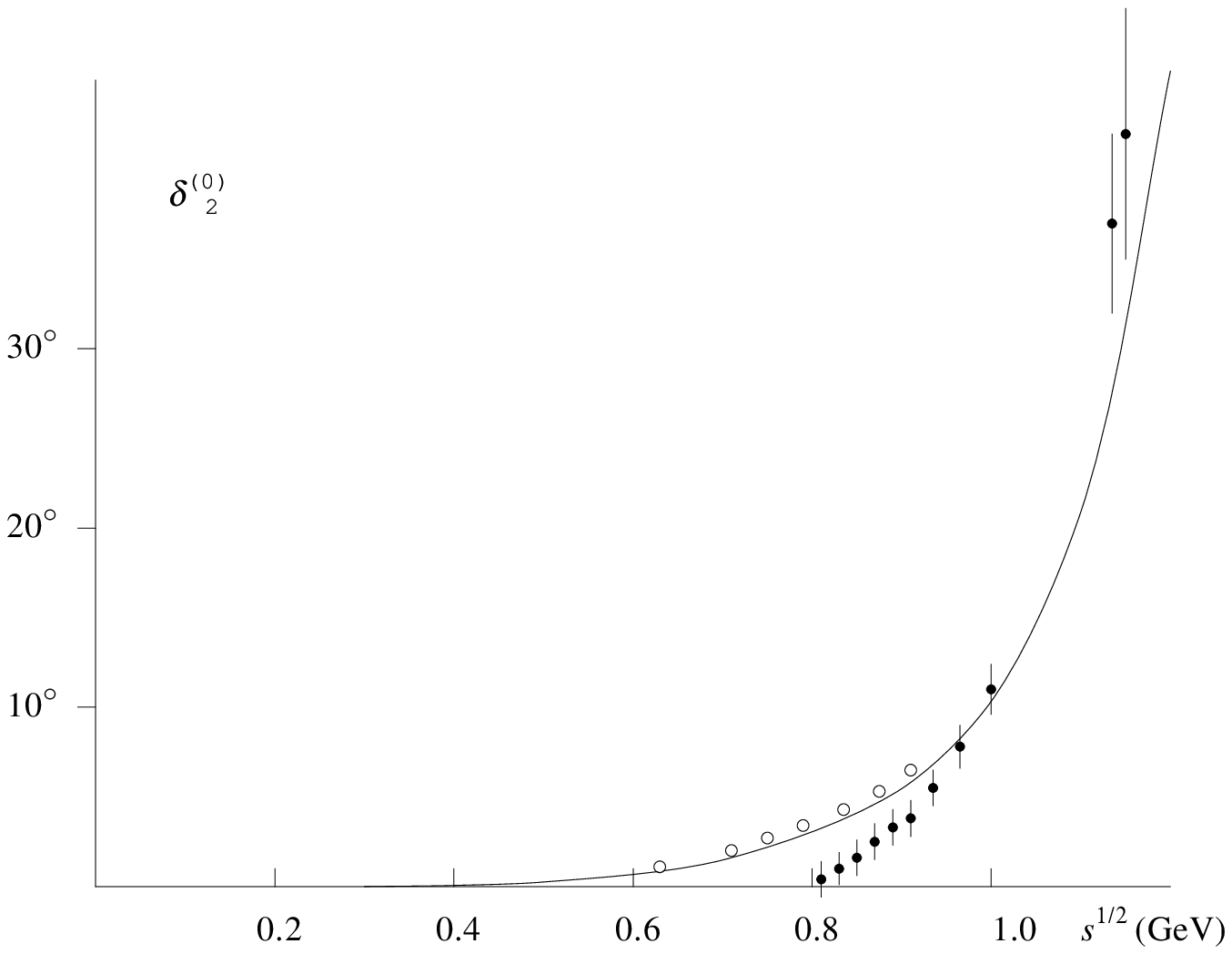}}} 
\setbox6=\vbox{\hsize 11truecm\captiontype\figurasc{Figure 3.3. }{
Fits to the  
$I=0$, $D$-wave phase shift. 
Also shown are the data points from solution~1 of Protopopescu et al.\ref{12} (black dots) and 
some data  of  Estabrooks and Martin\ref{11} (open circles).
}\hb} 
\centerline{\tightboxit{\box0}}
\bigskip
\centerline{\box6}
\medskip
}\endinsert
The inelasticity on the $f_2$ is taken from the Particle Data Tables. 
(3.9) corresponds to $a_2^{(0)}=(15\pm3.5)\times10^{-4}\,M^{-5}_\pi$ against CGL's value 
$(17.5\pm0.3)\times10^{-4}\,M^{-5}_\pi$.
For the D2 wave,\fnote{(3.10) corrects a mistake in the corresponding wave in ref.~6.}
$$\eqalign{
\cot\delta_2^{(2)}(s)=&\,\dfrac{s^{1/2}}{2k^5}\,
\dfrac{M_{\pi}^4s}{4(M_{\pi}^2+\deltav^2)-s}\,
\left\{B_0+B_1\dfrac{\sqrt{s}-\sqrt{s_0-s}}{\sqrt{s}+\sqrt{s_0-s}}\right\};\cr
s_0^{1/2}=1.43\;\gev;\quad
\quad B_0=&\,(2.33\pm0.17)\times10^3,
\quad B_1=(-0.39\pm0.75)\times10^3,\quad \deltav=90\pm11\;\mev.\cr
}
\equn{(3.10)}$$
Now $a_2^{(2)}=(1.6\pm0.4)\times10^{-4}\,M^{-5}_\pi$ [CGL's value: 
$(1.7\pm0.13)\times10^{-4}\,M^{-5}_\pi$].

The D phases are depicted in Figs.~3.3, 3.4.

Finally, for the F wave we write a background plus a Breit--Wigner. 
The background is obtained fitting low energy, the resonance is the $\rho_3$
 with its properties taken from the Particle Data Tables:
$$\eqalign{
\imag \hat{f}_3(s)=&\,\dfrac{1}{1+\cot^2\delta_3}+\left(\dfrac{k}{k(M_{\rho_3})}\right)^{14}
{\rm BR}\dfrac{M^2_{\rho_3}\gammav^2}{(s-M^2_{\rho_3})^2+M^2_{\rho_3}\gammav^2};\cr
\cot\delta_3(s)=&\,\dfrac{s^{1/2}}{2k^7}\,M^6_\pi\,
\left\{B_0+B_1\dfrac{\sqrt{s}-\sqrt{s_0-s}}{\sqrt{s}+\sqrt{s_0-s}}\right\};
\quad s_0^{1/2}=1.5\;\gev\cr
 M_{\rho_3}=&\,1.69\;\gev,
\quad \gammav=0.161\;\gev,\quad {\rm BR}=0.24;\cr
B_0=&\,(1.07\pm0.03)\times10^5,\quad B_1=(1.35\pm0.03)\times10^5.
\cr}
\equn{(3.11)}$$
Here $a_3=(7.0\pm0.8)\times10^{-5}\,M^{-7}_\pi$; the value reported in CGL is 
$(5.6\pm0.2)\times10^{-5}\,M^{-7}_\pi$.

\topinsert{
\setbox0=\vbox{\hsize13.truecm{\epsfxsize 10.8truecm\epsfbox{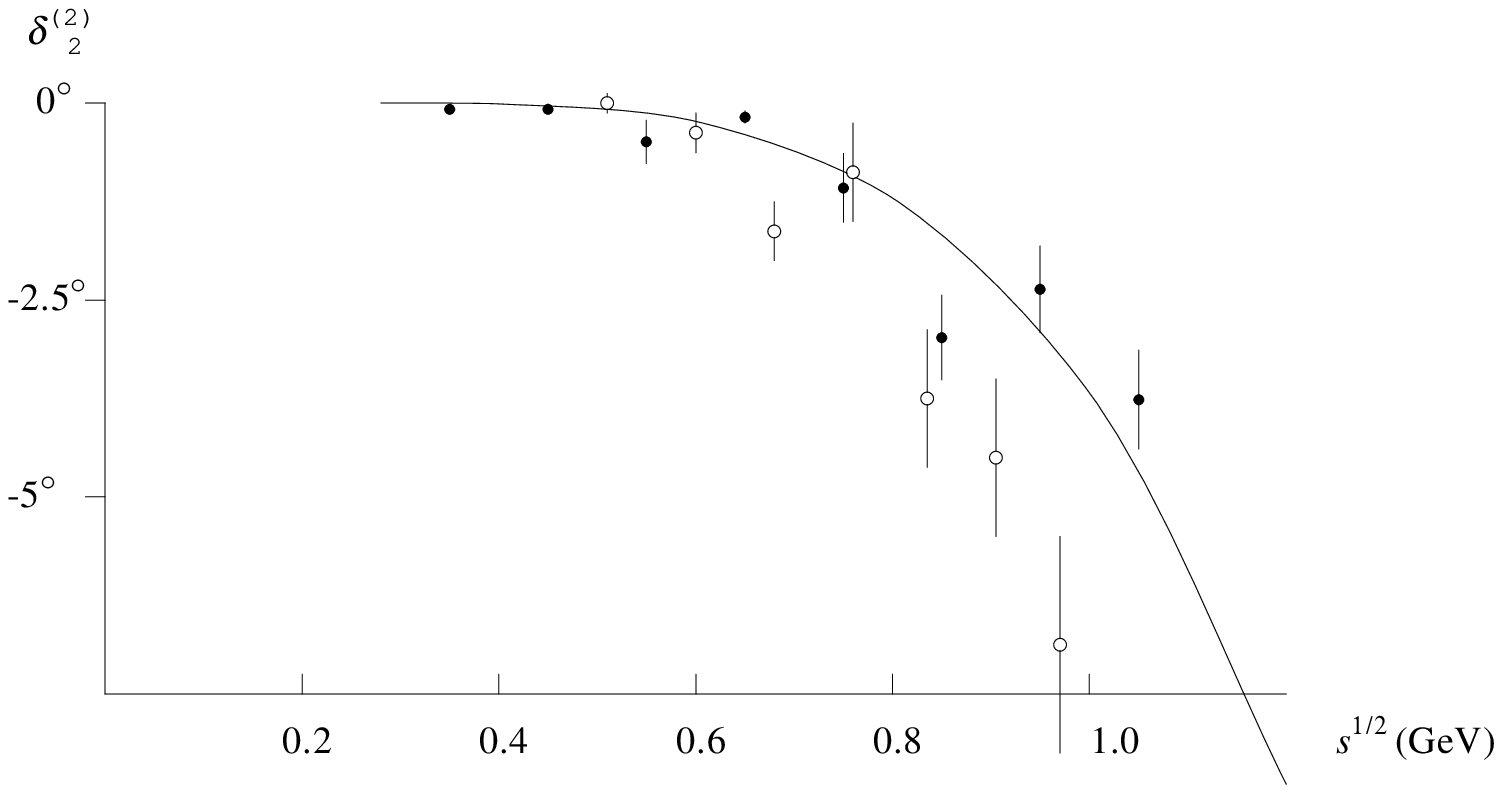}}} 
\setbox6=\vbox{\hsize 11truecm\captiontype\figurasc{Figure 3.4. }{
Fits to the  
$I=2$, $D$-wave phase shift. 
Also shown are the data points of Losty et al. (open circles) and 
 from solution~A of Hoogland et al. (black dots), refs.~13.
}\hb} 
\centerline{\tightboxit{\box0}}
\bigskip
\centerline{\box6}
\medskip
}\endinsert

\brochuresubsection{3.4. High energy: the Regge picture}

\noindent
As we will discuss in \sect~5, the experimental phase shift
 analyses become unreliable as soon as the inelasticity is large; 
for $\pi\pi$ scattering, this occurs at and above $E\sim1.4$ \gev. 
Fortunately, 
Regge pole theory provides  an input for high energy scattering; 
we will now briefly describe  those of its features that are of interest to us. 
Before starting with the details, however, it is 
perhaps worth while to remark that Regge theory is as much part of QCD as,
 say, chiral perturbation theory; in fact, Regge theory is probably of more general 
validity than QCD. By using Regge formulas we are thus not 
introducing extra assumptions. 
The only debatable point is when is Regge theory applicable; 
QCD only specifies $s\gg\lambdav^2$, $s\gg|t|$. 
Fortunately, factorization allows us to relate 
$\pi\pi$ to $\pi N$ and $NN$ cross sections. 
From this, and the fact that Regge formulas and experimental cross sections 
for $\pi\pi$ scattering agree (within errors) around $s^{1/2}=1.4\;\gev$, 
as shown in \figs~3.6-3.8 below,  
we will conclude that Regge formulas are applicable at and above these energies; 
specifically, we will use them above $E=1.42\,\gev$.
 We now turn to a brief discussion of the details.

Consider the collision of two hadrons, $A+B\to A+B$. 
According to Regge theory, the high energy scattering amplitude, 
at fixed $t$ and large $s$, 
is governed by the exchange of complex, composite objects (known as {\sl Regge poles}) 
related to the resonances that couple to the $t$ channel. 
Thus, for isospin 1 in the $t$ channel, high energy scattering
 is dominated by the exchange of a ``Reggeized" 
$\rho$ resonance.
 If no quantum number is exchanged, we say that the corresponding Regge pole is the vacuum, or
Pomeranchuk Regge pole; this name is often shortened to {\sl Pomeron}. 
In a QCD picture, the Pomeron (for example) will be associated with the exchange of 
a gluon ladder between two partons in particles $A$, $B$ (\fig~3.5). 
The corresponding formalism has been developed by Gribov, Lipatov and other Russian physicists in 
the 1970s, and is related to the so-called Altarelli--Parisi, or DGLAP mechanism 
in deep inelastic scattering.\ref{13}

\topinsert{
\setbox0=\vbox{\hsize 7truecm\epsfxsize 5.2truecm\epsfbox{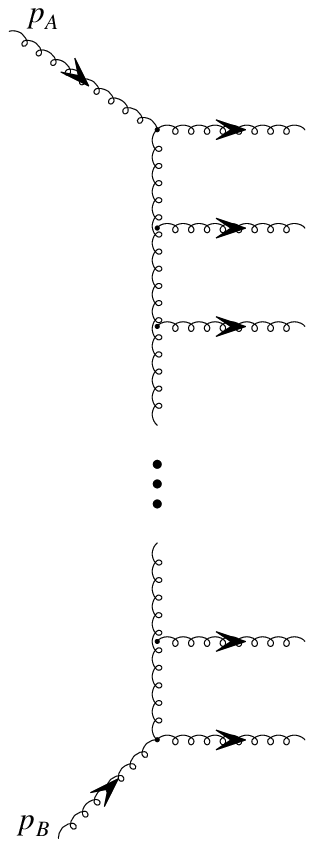}\hfil}
\setbox1=\vbox{\hsize 6cm\captiontype\figurasc{Figure 3.5. }
 {Cut Pomeron ladder exchanged between the partons $p_A$ and $p_B$ in 
hadrons $A$, $B$. The emitted gluons will materialize into a shower of particles. 
The cross section is proportional to the square of the cut ladder.}\hb
\vskip1.1cm
\phantom{x}}
\centerline{\box0\hfil\box1}
\vskip-0.5truecm
}\endinsert

One of the useful properties of Regge theory is  {\sl factorization};\ref{13}  
 it can be proved from general properties of Regge theory.\fnote{In potential theory 
the proof can be made mathematically rigorous; in relativistic theory, it follows 
from extended unitarity or, in QCD, in the DGLAP
formalism,  as is intuitively obvious from \fig~3.3.}
Factorization  states that, for example, the 
imaginary part of the scattering amplitude $F_{A+B\to A+B}(s,t)$
can be written as a product
$$\imag F_{A+B\to A+B}(s,t)\simeqsub_{{s\to\infty}\atop{t\,{\rm fixed}}}
f_A(t)f_B(t)(s/\hat{s})^{\alpha_R(t)}.
\equn{(3.12)}$$ 
Here $\hat{s}$ is a constant, usually taken to be $1\,\gev^2$ (we will do so here); 
the functions $f_A,\,f_B$ depend on the corresponding particles (if we had  
external currents, also on their virtuality),  
but the power $(s/\hat{s})^{\alpha_R(t)}$ is universal and
 depends only on the quantum numbers exchanged in 
channel $t$. 
The exponent $\alpha_R(t)$ is  the Regge trajectory associated to the 
quantum numbers in channel $t$ and, for small $t$, 
may be considered linear:
$$\alpha_R(t)\simeqsub_{t\sim0}\alpha_R(0)+\alpha'_Rt.
\equn{(3.13)}$$
For the $\rho$ and Pomeron pole, fits to high energy processes give
$$\eqalign{
\alpha_\rho(0)=&\,0.52\pm0.02,\quad\alpha'_\rho=
1.01\, {\gev}^{-2}\cr
\alpha_P(0)=&\,1,\quad\alpha'_P=0.11\pm0.03\, {\gev}^{-2}.\cr
}
\equn{(3.14)}$$
The 
Regge parameters taken here are essentially those in the global fit {\sl 1a} of Rarita et al.\ref{15}; 
for $\alpha_\rho(0)$, 
however, we take the value $0.52\pm0.02$ 
  which is more consistent with recent determinations
 based on deep inelastic scattering.\fnote{Fits
 to deep inelastic scattering processes, and references to  
previous literature, may be found in the book by FJY in ref.~14}
The results depend very little on this.

\topinsert{
\setbox3=\vbox{\hsize 11.5truecm\epsfxsize 10truecm\epsfbox{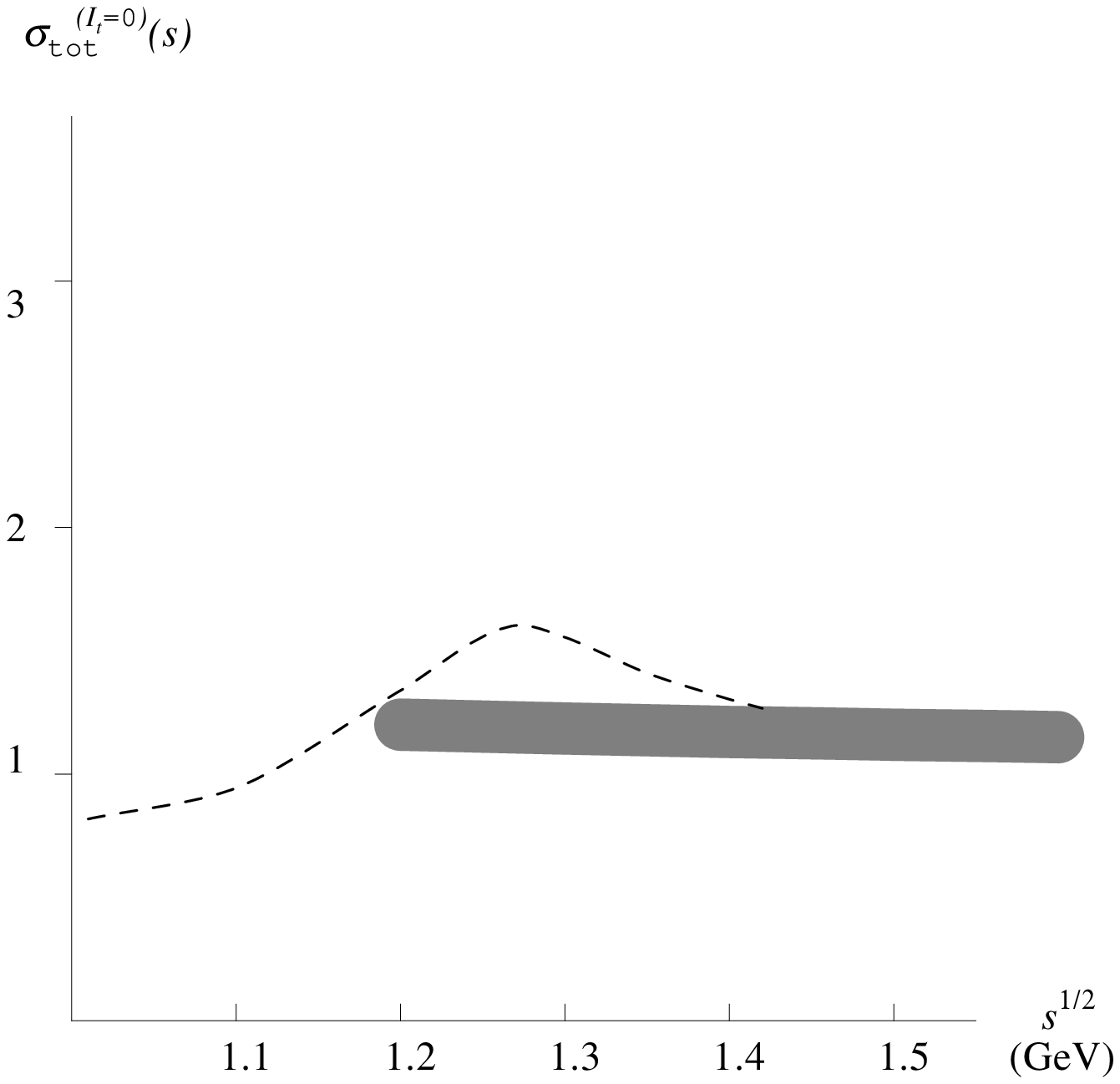}\hfil}
\setbox5=\vbox{\hsize 5truecm\captiontype\figurasc{Figure 3.6. }
 {The average cross section $\tfrac{1}{3}[2\sigma_{\pi^0\pi^+}+\sigma_{\pi^0\pi^0}]$, 
which is pure $I_t=0$,  
 arbitrarily normalized. 
Broken line: experimental cross section. Note that the bump here, as 
the larger bumps in 
\figs~3.7, 3.8,  
is due to the coincidence of two resonances, $f_0(1270),\,f_2(1370)$, 
mostly elastic, around $s^{1/2}\sim 1.3\,\gev$. 
Thick gray line: Regge formula (3.17a). The thickness of
 the line   covers the error in the theoretical value of the 
Regge residue.}}
\line{\box3\hfil\box5}
\bigskip
}\endinsert

Let us consider the imaginary part of the $\pi N$ or $NN$ scattering amplitudes (here 
by $NN$ we also understand $\bar{N}N$).
We have, 
$$\eqalign{
\imag F^{(I_t)}_{NN}(s,t)\simeq &\,\left[f^{(I_t)}_N(t)\right]^2(s/\hat{s})^{\alpha_R(t)},\quad
\imag F^{(I_t)}_{\pi N}(s,t)\simeq\,f^{(I_t)}_\pi(t)f^{(I_t)}_N(t)(s/\hat{s})^{\alpha_R(t)}.\cr
}
\equn{(3.15a)}$$
For $I_t=1$, $R=\rho$; for $I_t=0$, $R=P$ 
(the Pomeron).
Therefore, using factorization, we find
$$
\imag F^{(I_t)}_{\pi \pi}(s,t)\simeq \left[f^{(I_t)}_\pi(t)\right]^2(s/\hat{s})^{\alpha_R(t)}.
\equn{(3.15b)}$$
The functions $f^{(I_t)}_i(t)$ 
depend exponentially on $t$ for small $t$ and may be written, approximately, as\fnote{Consistency 
requires a more complicated form for the residue functions $f^{(I_t)}_i(t)$; 
see refs.~4, 15. For 
 the small values of $t$ in which we are interested, our expressions are sufficiently accurate.}
$$f^{(I_t=0)}_i(t)=\sigma_i(P)\ee^{bt},\quad
f^{(I_t=1)}_i(t)=\sigma_i(\rho)\dfrac{1+\alpha_\rho(t)}{1+\alpha_\rho(0)}\,
\left[(1+1.48)\ee^{bt}-1.48\right];
\qquad b=(2.38\pm0.20)\;{\gev}^{-2}.
\equn{(3.16)}$$
The exponent $b$ appears to be the same for rho,  Pomeron and $P'$, within errors.\ref{14}

From (3.15) we can deduce the relations among the cross sections
$$\dfrac{\sigma_{\pi\pi\to{\rm all}}}{\sigma_{\pi N\to{\rm all}}}=
\dfrac{\sigma_{\pi N\to{\rm all}}}{\sigma_{N N\to{\rm all}}},
$$
and from  these relations one can obtain the parameters $\sigma_\pi$ in (3.16) in terms of the 
known $\pi N$ and $NN$ cross sections. Using this, we can 
write explicit formulas for 
$\pi\pi$ scattering with exchange of isospin $I_t=0$ in the $t$ 
channel:
$$ 
\imag F^{(I_t=0)}(s,t)\simeqsub_{{s\to\infty}\atop{t\,{\rm fixed}}}
\left\{1+0.24\sqrt{\dfrac{\hat{s}}{s}}\right\}
\sigma_\pi(P)\ee^{bt}(s/\hat{s})^{\alpha_P(0)+\alpha'_P t},
\equn{(3.17a)}$$
and we have added empirically the subleading contribution, proportional to $\sqrt{\hat{s}/s}$, 
of the so-called $P'$ pole (associated with the $f_2$ resonance) that is 
necessary at the lowest energy range (see \fig~3.6). 
For $I_t=1$,
$$\eqalign{
\imag F^{(I_t=1)}_{\pi\pi\to\pi\pi}(s,t)\simeqsub_{{s\to\infty}\atop{t\,{\rm fixed}}}&\,
\imag F^{(\rho)}(s,t)+\imag F^{(\rm Bk)}(s,t),\cr
\imag F^{(\rho)}(s,t)=&\,\sigma_\pi(\rho)\dfrac{1+\alpha_\rho(t)}{1+\alpha_\rho(0)}\,
\left[(1+1.48)\ee^{bt}-1.48\right](s/\hat{s})^{\alpha_\rho(0)+\alpha'_\rho t},\cr
\imag F^{(\rm Bk)}(s,t)=&\,(0.4\pm0.1)\left(\dfrac{\hat{s}}{s}\right)^{1/2}
\imag F^{(\rho)}(s,t).\cr
}
\equn{(3.17b)}$$
We have added a background (Bk) contribution to the isospin 1 amplitude; 
this should be considered purely empirical and is 
adjusted so that the asymptotic formula joins 
smoothly the  experimental amplitude at low energy, within errors; see \fig~3.7.

\topinsert{
\setbox4=\vbox{\hsize 11.5truecm\epsfxsize 10truecm\epsfbox{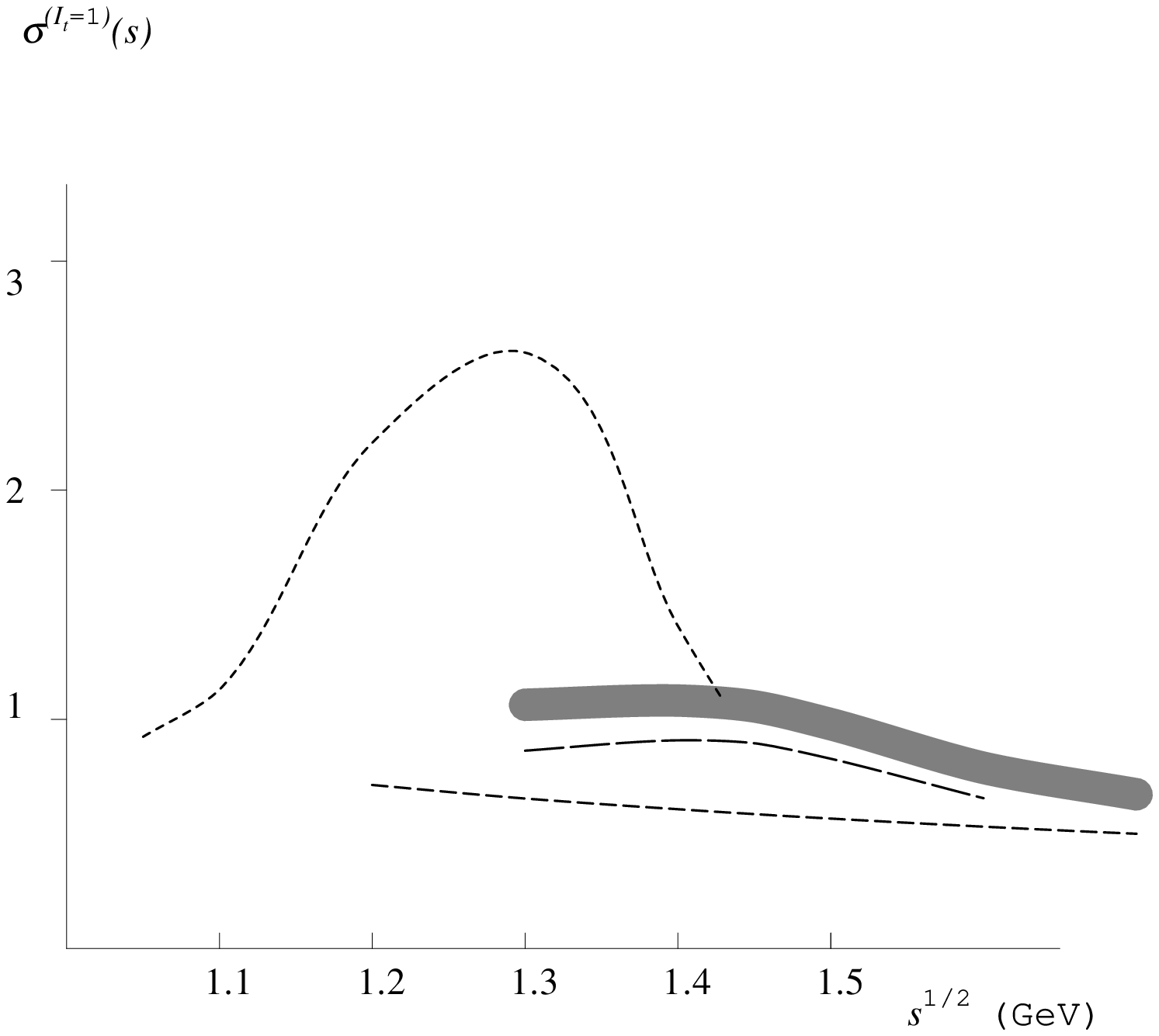}\hfil}
\setbox6=\vbox{\hsize 5truecm\captiontype\figurasc{Figure 3.7. }
 {The  cross section $\sigma^{(I_t=1)}$, 
for isospin 1 in the $t$ channel, arbitrarily normalized. 
The dotted line is experiment; the short-dashed line  the 
rho exchange Regge theory. The long-dashed line 
is obtained by adding to this the 
$\rho(1450)$ contribution. 
Finally, the thick gray line includes also the background Regge 
piece. The thickness of this line is equal to  the error 
due {\sl only} to the rho Regge residue (the total error for the 
full theoretical formula used in the text, that includes errors due to 
$\rho(1450)$ and the background Regge piece, is some 20\% larger).}}
\line{\box4\hfil\box6}
\bigskip
}\endinsert

From (3.16) and the known cross sections for $\pi N$, $NN$ scattering 
we have
$$\sigma_\pi(P)=3.0\pm0.3;\quad \sigma_\pi(\rho)=0.84\pm0.10
\equn{(3.17c)}$$
where the errors are obtained by considering the dispersion of the 
values of the parameters in ref.~15, and increasing the result by 50\%, 
which should cover amply the uncertainty on the point where one joins experimental and 
asymptotic formulas (that here we have taken to be 1.42 \gev) 
as well as errors in the parameters we have taken fixed.

It is important to note that the Regge parameters in the fit of
Rarita et al.  are obtained by global fit to $\pi N$, $NN$ and $\bar{N}N$ data for small
momentum transfer  and for c.m. kinetic energies in the region between 1 GeV and 6 GeV
(approximately),  which is the region of interest for us here as the contribution to the various
integrals  above this energy is negligible. The results of  Rarita et al. are still the best
available  as indeed there  are essentially no new data in that energy range. 
We have, on the other hand, verified that the cross sections are compatible with the 
corresponding values as given in the more recent editions of the Particle Data Tables.\ref{16}

We will treat the errors in the various Regge parameters as uncorrelated. 
In fact, the leading Regge amplitudes (Pomeron and rho) are 
uncorrelated; there is some correlation with, respectively, 
the $I_t=2$ exchange (see below) and the  Bk piece for $I_t=1$ exchange, 
because they have been fixed by fitting the sum to the pion cross sections. 
Since this only affects subleading pieces this would only have a minute influence in the 
results (in fact, they would slightly decrease the overall error due to the 
Regge contributions), and anyway the variations are substantially
smaller than
 the 50\%  increase in the errors of the
 Regge residues with which we have made our
evaluations.

For each individual process $\pi^0\pi^+$, $\pi^0\pi^0$, 
we have to incorporate 
the amplitude for exchange of isospin $I_t=2$ in the $t$ 
channel, which would be due to double rho exchange. 
This cannot be obtained from factorization, since $\pi N$ or $NN$ do 
not contain such amplitude. 
We use an empirical formula,
$$\imag F^{(I_t=2)}(s,t)= C_2\ee^{-bt}
\left[\imag F^{(\rho)}(s,t)\right]^2\left(\dfrac{\hat{s}}{s}\right),
\quad C_2=0.8\pm0.2,
\equn{(3.18)}$$
and we have obtained the constant $C_2$ by fitting the difference between the 
experimental $\pi^0\pi^0$ and  $\pi^0\pi^+$ total cross sections at $s^{1/2}=1.42\,\gev$, 
and the Pomeron plus $P'$ values; 
see \fig~3.8. 

The dependence of our results on $\imag F^{({\rm Bk})}$, $\imag F^{(I_t=2)}$ 
is very slight (for the second, with the exception of the $b_2^{(I)}$).

\topinsert{
\setbox4=\vbox{\hsize 11.5truecm\epsfxsize 10truecm\epsfbox{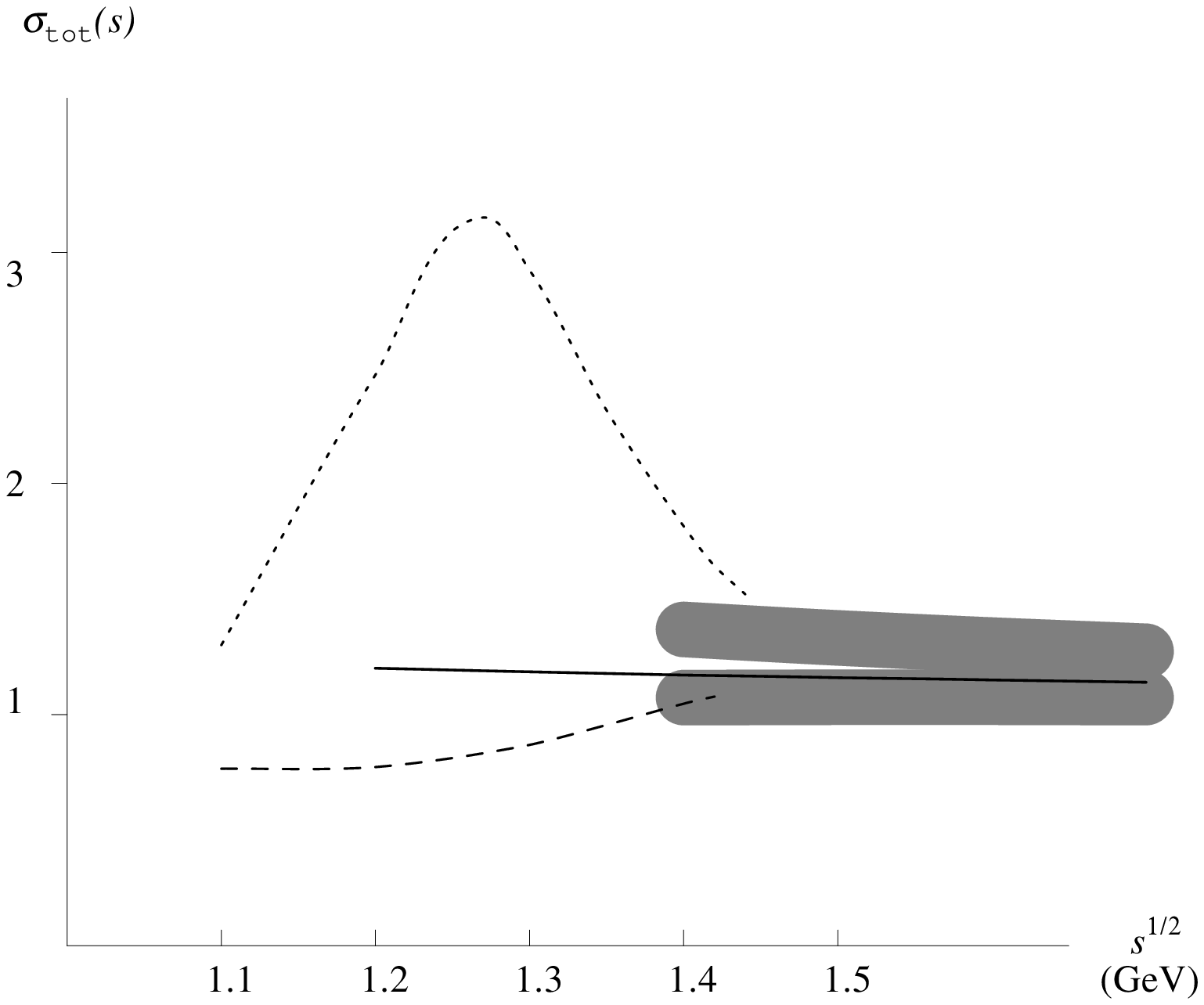}\hfil}
\setbox6=\vbox{\hsize 10.4truecm\captiontype\figurasc{Figure 3.8. }
 {The  cross sections $\sigma(\pi^0\pi^+)$ (dashed line),  $\sigma(\pi^0\pi^0)$ (dotted line), 
and the Pomeron plus  $P'$ (continuous line). 
The thick gray bands are obtained including the $I=2$ 
exchange contributions. Their thickness 
corresponds only to  the error of the 
Pomeron piece.}}
\centerline{\box4}
\bigskip
\centerline{\box6}
\bigskip
}\endinsert

We now add a few words on the matter of when one may apply  formulas like (3.17,\/18).
From the QCD, DGLAP version of the Pomeron, 
we expect the following pattern to occur: in the region $|t|\ll s$, $s\gg\lambdav^2$ (with
$\lambdav\sim0.3\,\gev$ the QCD 
parameter) the ladder exchange mechanism will start to dominate the collision $A+B$. 
We then will have the onset of the Regge regime with, at the same time, 
a large increase of inelasticity and a smoothing of the total cross 
section according to the behaviour (3.17).  

For $\pi N$, $NN$
scattering this occurs as soon as one is
 beyond the region of elastic resonances; in fact
  (as can be seen in the cross section summaries in the 
Particle Data Tables) as soon as the kinetic energy or lab momentum is above 
$1$ to $1.2$ \gev. For $\pi\pi$ 
we thus expect the Regge description to be valid for 
the corresponding energies, that is to say, for  $s^{1/2}\gsim 1.4\,\gev$. Indeed, 
 around $s^{1/2}\sim 1.4\,\gev$ it is still possible to calculate the 
$\pi\pi$ scattering 
amplitudes reliably from 
experimental phase shifts and indeed they agree, within a 10\%, with 
the Regge expressions in  
the $\pi^0\pi$ cases; 
see Figs.~3.6,~3.8. 
Moreover, the experimental inelasticity for $\pi\pi$ around 1.4 \gev,  $\sim20\%$, 
also agrees 
with the 
value of the inelasticity measured at the same energies for $\pi N$ or $NN$ scattering.

For the $I_t=1$ amplitude, and because it is a difference between large 
amplitudes, the influence of resonances may be expected to extend to higher energies. 
Indeed, we see in Fig.~3.7 that agreement between experiment 
and the Regge expression (within errors) around 1.4 \gev\ requires adding
 the resonance 
$\rho(1450)$, 
as in \equn{(3.8)}. We will do so in our calculations. 
Thus, for all $\pi\pi$ amplitudes we will assume the Regge formula 
(eventually adding the $\rho(1450)$ contribution) to be valid for 
 $s^{1/2}\geq 1.42\,\gev$.

As is clear from this minireview, the reliability of the Regge calculation of 
high energy pion-pion scattering cannot go beyond an accuracy of $\sim 10 \%$, even 
for small $t$. 
The deviations off simple Regge behaviour are expected to be much larger for large $|t|$, 
because the counting rules of QCD imply a totally different behaviour for fixed $t/s$. 
This is one of the problems involved in using e.g. the Roy equations that require 
integration up to $-t\sim s\sim 1.7\,\gev^2$, where the Regge picture fails completely 
(we expect instead the Brodsky--Farrar behaviour, $\sigma_{{\rm fixed}\,\cos\theta}\sim s^{-5}$). 
However, for forward dispersion relations or the Froissart--Gribov representation 
 we will work only for $t=0$ or $t=4M_{\pi}^2$ for 
which  the largest variation, that of $\ee^{bt}$, is still small, since 
$b\times(t=4M_{\pi}^2)\simeq0.19$. So we expect no large error 
due to departure off linearity\fnote{In the case of the 
rho trajectory, exact linearity would imply 
$\alpha'_\rho(0)=1/2(M^2_\rho-M^2_\pi)\simeq0.87\,{\gev}^{-2}$, 
not far from the value $1.01\,{\gev}^{-2}$ that the actual fits give, and which we have used here.} 
for the exponent in $f^{(I_t)}_i(t)$ or for the Regge trajectories, 
$\alpha_R(t)$.

\brochuresection{4. Olsson's sum rule and the Froissart--Gribov calculation
 of $a_1$, $b_1$, $a_2^{(I)}$, $b_2^{(I)}$
}
\vskip-0.5truecm
\brochuresubsection{4.1. The Olsson sum rule}
The Olsson sum rule is simply a forward dispersion relation for the amplitude
$F^{(I_t=1)}$ with isospin 1 in the $t$ 
channel, evaluated at threshold. 
Expressing $F^{(I_t=1)}(4M^2_\pi,0)$  in terms of the scattering lengths,  
this reads
$$2a_0^{(0)}-5a_0^{(2)}=D_{\rm O},\quad
D_{\rm O}=3M_\pi\int_{4M_\pi^2}^\infty \dd s\,
\dfrac{\imag F^{(I_t=1)}(s,0)}{s(s-4M_\pi^2)}.
\equn{(4.1)}$$
In terms of isospin in the $s$ channel,
$$F^{(I_t=1)}(s,t)=\tfrac{1}{3}F^{(I_s=0)}(s,t)+\tfrac{1}{2}F^{(I_s=1)}(s,t)-
\tfrac{5}{6}F^{(I_s=2)}(s,t);
\equn{(4.2)}$$
the $F^{(I_s)}$ are normalized by
$$\imag F^{(I_s)}(s,t)=2\dfrac{2s^{1/2}}{\pi k}\sum_l(2l+1)P_l(\cos\theta)\imag\hat{f}_l^{(I_s)}(s).
$$ 
Substituting in the right hand side above the  S, P phases of CGL up to $0.82\,\gev$, the 
phases (as given in the parametrizations of \subsects~3.2,3) for the same at intermediate energies 
($0.82\leq E\leq1.42\,\gev$), the D, F phases from (3.9-11), 
the tail of the $\rho(1450)$ resonance between 1.42 
and 1.6 \gev, and the $\rho$ plus background Regge parameters of 
\subsect~3.4 we find, for $2a_0^{(0)}-5a_0^{(2)}$ in units of $M_\pi$,
\smallskip
$$\matrix{
{\rm CGL},\;{\rm direct}&{\rm CGL},\;{\rm dispersive}&\cr
&0.400\pm0.007&\quad\hbox{[CGL S, P, $\;s^{1/2}\leq0.82\;\gev$]}\cr
&             0.146\pm0.004&\quad\hbox{[Rest, $s^{1/2}\leq 1.42\,\gev$
 (incl., D, F below 0.82 \gev)]}\cr
&             0.073\pm0.010&\quad\hbox{[Regge, $\rho\; \;s^{1/2}\geq1.42\;\gev$]}\cr
&             0.010\pm0.003&\quad\hbox{[Regge, Bk;$\;s^{1/2}\geq1.42\;\gev$]}\cr
&             0.005\pm0.001&\quad \rho(1450), 1.42\leq s^{1/2}\leq1.6\,\gev\cr
0.663\pm0.007&\phantom{\Bigg{|}}             0.635\pm0.014&\quad\hbox{[Total, disp.]}\cr
}
\equn{(4.3)}$$

By ``direct" we mean the value of the corresponding quantity (in our case, $2a_0^{(0)}-5a_0^{(2)}$) 
as given in CGL. By ``dispersive" we understand that we 
have used the dispersive formula, $D_{\rm O}$ in (4.1), to calculate the 
same quantity. The ``Rest"  are the contributions  of the 
D, F waves below $1.42\,\gev$, plus the S, P waves between 
0.82 and 1.42 \gev. Of this ``Rest",  the largest 
contribution comes from the 
D0 and P waves.

The error in the CGL~S, P piece below 0.82 \gev\ we obtain by varying the $A,\,B,\,C,\,D$ parameters 
in (3.1) according to the formulas given by ACGL (Appendix). It is almost  
identical to the error given for the whole of 
the ``direct" quantity itself.
We will discuss in some detail the discrepancy between
the ``direct"  and ``dispersive" determinations of this quantity $2a_0^{(0)}-5a_0^{(2)}$ 
as the situation for the other $a$s and $b$s to be 
considered below will be very similar.

The reason the analysis of the discrepancy is not straightforward is that both 
determinations are strongly correlated, as they both 
depend on the same parameters. 
The ``direct" determination is obtained from the parameters $A,\,B,\,C,\,D$ in CGL
 (as given in Eq.~(3.1) here), which describe in particular the S waves. 
So we should really write
$$[2a_0^{(0)}-5a_0^{(2)}]^{\hbox{``direct"}}_{A,B,C,D}.$$
The integrals in the dispersive determination contain
 the contributions of the S, P waves up to $0.82$ GeV, which 
are given by the same $A,\,B,\,C,\,D$, so one also has
$$[2a_0^{(0)}-5a_0^{(2)}]^{\hbox{``dispersive, S, P CGL"}}_{A,B,C,D}.$$
Now, it is clear that if we try to change the parameters
 $A,\,B,\,C,\,D$  in the ``direct" determination to, for example,
decrease the value of $a_0^{(0)}$ 
to bring this closer to the dispersive value, the same change in the 
 $A,\,B,\,C,\,D$  will affect the integral over the S0 wave in 
$[2a_0^{(0)}-5a_0^{(2)}]^{{{\rm``dispersive,\, S,\, P\, CG"}}}_{A,B,C,D},$ making this
smaller
 and therefore nullifying to
a large extent the improvement.

What one has to do to solve this problem is to consider the difference
$\Delta=2a_0^{(0)}-5a_0^{(2)}-D_{\rm O}$
and vary here the parameters $A,\,B,\,C,\,D$. 
Then we find the 
 value 
$$\Delta=0.027\pm0.011:$$
 that is to say, a $2.5\,\sigma$ discrepancy. 

This procedure will also be followed for the Froissart-Gribov sum rules,
 where the correlation in the CGL analysis 
is transmitted in part by the common chiral perturbation theory parameters $\bar{l}_i$.
  (We will discuss more
about errors in Subsects.~4.4 and 5.1).

\brochuresubsection{4.2. The Froissart--Gribov representation:  $a_1$, $b_1$}

\noindent
By projecting the dispersion relation (2.1) (or a derivative with respect to $t$ of it) 
over the $l$th partial wave in the $t$ channel, at $t=4M^2_\pi$, one finds the Froissart--Gribov 
representation
$$\eqalign{a_l=&\,\dfrac{\sqrt{\pi}\,\gammav(l+1)}{4M_{\pi}\gammav(l+3/2)}
\int_{4M_{\pi}^2}^\infty\dd s\,\dfrac{\imag F(s,4M_{\pi}^2)}{s^{l+1}},\cr
b_l=&\,\dfrac{\sqrt{\pi}\,\gammav(l+1)}{2M_{\pi}\gammav(l+3/2)}
\int_{4M_{\pi}^2}^\infty\dd s\,
\left\{\dfrac{4\imag F'_{\cos\theta}(s,4M_{\pi}^2)}{(s-4M_{\pi}^2)s^{l+1}}-
\dfrac{(l+1)\imag F(s,4M_{\pi}^2)}{s^{l+2}}\right\},\cr
}
\equn{(4.4)}$$
$\imag F'_{\cos\theta}\equiv (\partial/\partial\cos\theta_s)\imag F$.
For amplitudes with fixed isospin in the $t$ channel, an extra factor 2 
(due to identity of particles) has to be added to the left hand side; 
so we have, for example,
$$2a_l^{(I=1)}=\dfrac{\sqrt{\pi}\,\gammav(l+1)}{4M_{\pi}\gammav(l+3/2)}
\int_{4M_{\pi}^2}^\infty\dd s\,\dfrac{\imag F^{(I_t=1)}(s,4M_{\pi}^2)}{s^{l+1}}.
$$

With the same type of calculation as for the Olsson sum rule, and 
with the same definitions, we now find, in units of $10^{-3}\,\times M_{\pi}^{-3}$,
\smallskip
$$\matrix{
a_1,\;{\rm CGL},\;{\rm direct}&{\rm CGL},\;{\rm Froissart-Gribov}&
\hbox{TY (St.)}&\hbox{TY (St.+Sys.)}\cr
&18.5\pm0.2\quad\hbox{[CGL S, P, $\;s^{1/2}\leq0.82\;\gev$]}&&\cr
&9.1\pm0.3\quad\hbox{[Rest, $s^{1/2}\leq 1.42\,\gev$]}&&\cr
&8.1\pm1.1\quad\hbox{[Regge, $\rho$]}&&\cr
&1.0\pm0.3\quad\hbox{[Regge, Bk]}&&\cr
&0.3\pm0.1\quad \rho(1450)&&\cr
37.9\pm0.5\phantom{\Bigg{|}}&37.1\pm1.3
\quad\hbox{[Total, Froissart--Gribov.]}&40.6\pm1.4&38.6\pm1.2.\cr
}
\equn{(4.5)}$$
Here, and for $b_1$, we profit from the existence of an independent determination of 
the P wave parameters, using the pion form factor data both in the timelike 
and in the spacelike regions, \ref{10} denoted by TY. 
From this we have chosen two values: from the fit taking into account only the statistical 
errors in the 
various data sets (St.), as in \equn{(3.4)} here; or 
taking also into account the systematic normalization errors (St. + Sys.), 
as in \equn{(3.5)}.

The distance between the direct evaluation and the one with the Froissart-Gribov calculation is 
now  $0.6\,\sigma$, and there is also acceptable overlap 
with the TY~(St.+Sys.) figure.

For the quantity $b_1$ we have,  in units of  $10^{-3}\,\times M_{\pi}^{-5}$,
\smallskip
$$\matrix{
b_1,\;{\rm CGL},\;{\rm direct}&{\rm CGL},\;{\rm Froissart-Gribov}&
\hbox{TY (St.)}&\hbox{TY (St. + Sys.)}\cr
&-0.92\pm0.05\quad\hbox{[CGL S, P, $\;s^{1/2}\leq0.82\;\gev$]}&&\cr
&1.02\pm0.04\quad\hbox{[Rest, $s^{1/2}\leq 1.42\,\gev$]}&&\cr
&5.33\pm0.86\quad\hbox{[Regge, $\rho$]}&&\cr
&0.55\pm0.16\quad\hbox{[Regge, Bk]}&&\cr
&0.01\pm0.0\quad \rho(1450)&&\cr
5.67\pm0.13&\phantom{\Bigg{|}}5.99\pm0.88
\quad\hbox{[Total, Froissart--Gribov.]}&4.18\pm0.43&4.47\pm0.29.\cr
}
\equn{(4.6)}$$
Here the Regge contribution is particularly important because the lower energy pieces cancel 
almost completely. 
The numbers labeled TY, as before, refer to what one obtains from  
the fit to the pion form factor. 
We remark that this last is a very robust determination in that it is obtained by
 fitting some 210 points from 
several independent experiments, is independent 
of high energy assumptions and it covers spacelike as well as timelike momenta: thus, the 
values of the threshold parameters are obtained by {\sl interpolation}, 
notoriously more stable than extrapolations.

There is no inconsistency between the ``direct" and Froissart--Gribov numbers for 
the CGL calculation, but they are both   too large by 
almost $4\,\sigma$ compared to even the more favorable 
value,  TY(St.+Sys.), following from the pion form factor.

\brochuresubsection{4.3. The Froissart--Gribov representation:  $a_2^{(I)}$,  
  $b_2^{(I)}$;
 $I=0,\,2$}

\noindent
We first 
calculate the two combinations of scattering lengths
 $a_{0+}=\tfrac{2}{3}[a_2^{(0)}-a_2^{(2)}]$ and 
$a_{00}= \tfrac{2}{3}[a_2^{(0)}+2a_2^{(2)}]$. 
They correspond to the $s-$channel amplitudes
$$F_{\pi^0\pi^+}=\tfrac{1}{2}F^{(I_s=1)}+\tfrac{1}{2}F^{(I_s=2)},\quad
F_{\pi^0\pi^0}=\tfrac{1}{3}F^{(I_s=0)}+\tfrac{2}{3}F^{(I_s=2)}.$$
The only important difference with the cases in the previous subsection is that 
the dominant high energy part is given now by the Pomeranchuk 
trajectory (instead of the rho) and its importance 
is small because the integrals converge faster.
We  find, in units of $10^{-4}\,\times M_{\pi}^{-5}$,
\smallskip
$$\matrix{
a_{0+},\;{\rm CGL},\;{\rm direct}&{\rm CGL},\;{\rm Froissart-Gribov}\cr
&8.43\pm0.09\quad\hbox{[CGL S, P, $\;s^{1/2}\leq0.82\;\gev$]}\cr
&1.84\pm0.05\quad\hbox{[Rest, $s^{1/2}\leq 1.42\,\gev$]}&\cr
&0.68\pm0.07\quad\hbox{[Regge, $I_t=0$]}\cr
&-0.06\pm0.02\quad\hbox{[Regge, $I_t=2$]}\cr
&0.04\pm0.01\quad [\rho(1450)&\cr
10.53\pm0.10\phantom{\Bigg{|}}&10.94\pm0.13
\quad\hbox{[Total, Froissart--Gribov.]}\cr
}
\equn{(4.7)}$$
In finding the error of the ``direct" value, $(10.53\pm0.10)\times10^{-4}\,M_{\pi}^{-5}$,
 it is important to take into account 
the strong correlations of the errors of the $a_2^{(0)},\,a_2^{(2)}$. 
To do this, we use Eq.~(14.4) in ACGL to calculate directly the quantity 
$a_{0+}$. 
The difference between the ``direct" and Froissart--Gribov values, 
with correlations taken into account, as we did in the 
case of the Olsson sum rule, is now 
$$0.38\pm0.09,$$
 so that
the discrepancy reaches the $4\,\sigma$ level.

In the same units,  $10^{-4}\,\times M_{\pi}^{-5}$, we have
\smallskip
$$\matrix{
a_{00},\;{\rm CGL},\;{\rm direct}&{\rm CGL},\;{\rm Froissart-Gribov}\cr
&11.73\pm0.32\quad\hbox{[CGL S, P, $\;s^{1/2}\leq0.82\;\gev$]}\cr
&1.91\pm0.04\quad\hbox{[Rest, $s^{1/2}\leq 1.42\,\gev$]}&\cr
&0.68\pm0.07\quad\hbox{[Regge, $I_t=0$]}\cr
&0.12\pm0.04\quad\hbox{[Regge, $I_t=2$]}\cr
13.94\pm0.32\phantom{\Bigg{|}}&14.44\pm0.33
\quad\hbox{[Total, Froissart--Gribov];}\cr
}
\equn{(4.8)}$$
we have also taken into account the correlations \`a la ACGL to evaluate the error of 
the ``direct" number. 
The difference between ``direct" and F.--G. values for CGL are, 
with correlations taken into account, of 
$$0.49\pm0.09,$$
 i.e., a $5\,\sigma$ discrepancy.

Finally, we present the results for  $b_{0+}=\tfrac{2}{3}[b_2^{(0)}-b_2^{(2)}]$ and 
$b_{00}= \tfrac{2}{3}[b_2^{(0)}+2b_2^{(2)}]$, both in units of $10^{-4}\,\times M_{\pi}^{-7}$:
\smallskip
$$\matrix{
b_{0+},\;{\rm CGL},\;{\rm direct}&{\rm CGL},\;{\rm Froissart-Gribov}\cr
&-0.331\pm0.015\quad\hbox{[CGL S, P, $\;s^{1/2}\leq0.82\;\gev$]}\cr
&0.04\pm0.00\quad\hbox{[Rest, $s^{1/2}\leq 1.42\,\gev$]}&\cr
&0.12\pm0.02\quad\hbox{[Regge, $I_t=0$]}\cr
&-0.05\pm0.02\quad\hbox{[Regge, $I_t=2$]}\cr
-0.189\pm0.016\phantom{\Bigg{|}}&-0.233\pm0.036
\quad\hbox{[Total, Froissart--Gribov.]}\cr
}
\equn{(4.9)}$$
The contribution of the resonance $\rho(1450)$ is now negligible. 
For the difference between the direct and Froissart--Gribov result we have
$$0.044\pm0.026,$$
that is to say, almost a $2\,\sigma$ discrepancy.
For $b_{00}$,
\smallskip
$$\matrix{
b_{00},\;{\rm CGL},\;{\rm direct}&{\rm CGL},\;{\rm Froissart-Gribov}\cr
&-6.90\pm0.22\quad\hbox{[CGL S,  $\;s^{1/2}\leq0.82\;\gev$]}\cr
&0.07\pm0.01\quad\hbox{[Rest, $s^{1/2}\leq 1.42\,\gev$]}&\cr
&0.12\pm0.02\quad\hbox{[Regge, $I_t=0$]}\cr
&0.10\pm0.05\quad\hbox{[Regge, $I_t=2$]}\cr
-6.72\pm0.22\phantom{\Bigg{|}}&-6.62\pm0.23
\quad\hbox{[Total, Froissart--Gribov.]}\cr
}
\equn{(4.10)}$$
For   $b_{00}$  the direct result and the one following from the 
Froissart--Gribov representation differ by $2\,\sigma$: 
$$0.10\pm0.05.$$ 
However,    
 one cannot take this or the discrepancy for $b_{0+}$  
 as seriously as in the  previous cases. 
This is so because of the large (relative) size of the contribution of the $I_t=2$ exchange 
piece, 
proportional to the derivative with respect to $t$ of  an expression we 
have  obtained purely empirically by fitting at $t=0$.

\brochuresubsection{4.4. How significant are the discrepancies?}

\noindent
In the present subsection we investigate whether the 
inconsistencies we have found can be eliminated (or to 
what extent they can be made less severe) by altering 
the non-CGL part of the dispersive, or Froissart--Gribov calculations. 
We will do so in two steps. 
First, we will consider what happens if we
alter the pieces labeled ``Rest" in (4.3) to (4.10); 
then we will address the question of what can be done at high energy ($s^{1/2}\geq1.42\,\gev$).
\smallskip
\noindent
4.4.1. {\sl The region between $0.82$ and $1.42$ \gev}
\medskip\noindent
We start with the first question, that we discuss in detail for the 
Olsson sum rule since the results for the Froissart--Gribov  
calculations are very similar. 
We then consider the following set of drastic modifications of 
our calculations: 
For the S0 wave, and $0.82\leq E\leq0.992\,\gev$ we may replace (3.2) by the CGL 
parametrization, (3.1). For the S0 wave and 
$0.992\leq E\leq1.42\,\gev$, where it is poorly known, we 
allow $\delta_0^{(0)}$ to vary between the two extreme 
values $\pi$ and $3\pi/2$. For the S2 wave, we multiply by 3 the 
errors given in (3.3). For the P wave, and $1\leq E\leq1.42\,\gev$, 
we change the elasticity of the $\rho(1450)$ resonance  by 50\% (up and down). 
For the D0 wave, that supplies the more important contribution to ``Rest", 
we consider the effect of taking the $f_2(1270)$ resonance to be purely elastic, 
or 30\% inelastic. 
The remaining contributions to ``Rest" are so small that 
we need not worry about them.
\goodbreak
The alterations just discussed are rather extreme; 
nevertheless, their effects are of no relevance. 
They produce the following extra errors (we give the central value of each term as well):
\smallskip
$$\matrix{
{\rm S0},\;0.82\leq s^{1/2}\leq0.992\,\gev:&\quad 0.026^{+0.0}_{-0.006}\cr
{\rm S0},\;0.992\leq s^{1/2}\leq1.42\,\gev:&\quad  0.018^{+0.005}_{-0.013} \cr
{\rm S2},\;0.82\leq s^{1/2}\leq1.42\,\gev:&\quad -0.022\pm0.004\cr
}
$$
\smallskip
$$\matrix{
{\rm P},\;1.0\leq s^{1/2}\leq1.42\,\gev:&\quad 0.024\pm0.005\cr
{\rm D0},\; s^{1/2}\leq1.42\,\gev:&\quad 0.055\pm0.001.\cr
}
$$
Including these increased errors  
we get that, for the Olsson sum rule, the result for the
 ``Rest" changes according to 
$$\hbox{``Rest"}:\quad 0.145\pm0.004\to0.145^{+0.009}_{-0.016},$$
and, for the whole dispersive result, we now get
$$\hbox{Total}:\quad 0.631\pm0.013\to0.631^{+0.015}_{-0.019},$$
i.e., practically no change at all in the upper error bar.
\smallskip
\noindent
4.4.2. {\sl The high energy region, $s^{1/2}\geq1.42$ \gev}
\medskip\noindent
Once we have verified that the inconsistencies between the CGL 
direct and dispersive calculations of low energy parameters cannot be due to 
the contributions of the intermediate energy region, 
we turn to the high energy ($s^{1/2}\geq1.42$) piece. 
Then, we relax the condition of factorization for the $\rho$ 
and Pomeron Regge residues (but we do not change the others). We 
 treat them now as free parameters, describing an {\sl effective} 
scattering amplitude, to see under which conditions 
one can reconcile the direct and 
Froissart--Gribov (or dispersive) 
evaluations for the scattering lengths and effective range, in the CGL-like analysis. 
Starting with the isospin 1 case, 
we thus write
$$\imag F^{(\rho)}_{\rm eff}(s,t)\simeqsub_{{s\to\infty}\atop{t\,{\rm fixed}}}
\lambda\,\sigma_\pi(\rho)\dfrac{1+\alpha_\rho(t)}{1+\alpha_\rho(0)}\,
\left[(1+1.48)\ee^{bt}-1.48\right](s/\hat{s})^{\alpha_\rho(0)+\alpha'_\rho t},
$$
that is to say, we modulate the $\rho$ amplitude in (3.17b) by the constant $\lambda$. 
We then fix $\sigma_\pi(\rho)=0.85$, and treat $\lambda$ as a free 
parameter. We then find that overlap 
between the direct and dispersive  determinations for the quantity 
$2a_0^{(0)}-5a_0^{(2)}$ involved in the Olsson sum rule
would require $\lambda=1.4$, which is  well outside expectations  and, 
 moreover, this  
spoils the overlap  for $a_1$, $b_1$, which become inconsistent
 at the  $2$ to $2.5\,\sigma$ level.

For the $a_2^{(I)}$ the situation is even more transparent. 
Consider for example the quantity $a_{00}$, 
Eq.~(4.8). Integrating only to 0.82, with the CGL phases, we find 
$11.73$, which is the bulk of the result.
 Even if the errors of what we call ``Rest"  were underestimated 
by a factor 3, and this ``Rest" would be 
1.79 (instead of 1.91), adding it one would get at least $13.52\pm0.33$ for 
the contribution below 1.42 GeV. The direct result, 
with the CGL  values of the $a_l^I$, is $13.94$. To get agreement, one  
would require the high energy, $E>1.42$ (Regge) estimate to be wrong by a factor 2, 
very difficult to believe. 
And it would be no good: the same Pomeron that contributes to  $a_{00}$ 
contributes to $a_{0+}$ and to the $b_{0+},\,b_{00}$.  
The disagreement would be shifted to the $b_{0+},\,b_{00}$, 
which would then be wrong by about $4\,\sigma$, and 
$a_{0+}$ would  still be wrong by almost $2\,\sigma$. 
As for the proverbial square peg in the round hole, 
trying to fit a corner only makes  others sick out more sharply.

\brochuresection{5 Discussion of the ACGL and CGL analyses}
\vskip-0.5truecm
\brochuresubsection{5.1. Possible cause of the distortion of the CGL solution} 

\noindent
In this section we try to ascertain the reasons for the 
troubles that seem to afflict  the CGL analysis. 
This is particularly important because, although 
ACGL or CGL did not verify the Froissart--Gribov relations, they {\sl did} check 
relations similar to  the 
Olsson sum rule. 
It follows that the reasons for the discrepancies must be due to the 
high energy 
input.
Here you have two regions: between 0.82 and 1.42 \gev\
(more or less) the inelasticity is low, and, as we have shown,  one can trust 
the experimental phase 
shifts. 
Even if they have systematic errors, these will likely not be 
large and they will just produce a slight fluctuation of the solution of the 
Roy equations, as we have shown explicitly in \subsect~4.4.1 that it occurs for our evaluations.

The difficult region, however, is for $s^{1/2}$ above 1.42 GeV. 
Between 1.42 and 2 \gev, CGL presumably use the phase shifts of ref.~11  
and, above 2 \gev, a Regge-type formula. 
We start the discussion with the region $1.42\leq s^{1/2}\leq2\,\gev$.
Here inelasticity is very high, and the  phase shifts and
inelasticity parameters cannot be determined reliably, at the 
level of accuracy required.\fnote{This unreliability is reflected, for 
example, in the Particle Data Tables (e.g., the edition of ref.~15),
 where no number is given for the 
branching ratios of $\pi\pi$ resonances with masses at or above 
1.2 GeV (with the exception of the $\rho_3(1690)$) 
and even the S0 phase around the $f_0(980)$ has a dubious status: 
this last due to the ambiguity caused by 
the inelastic $\bar{K}K$ channel. 
In fact, the resonances that appear in $\pi\pi$ production are not the 
same that one finds in $e^+e^-$, $\tau$ decay or $J/\psi$ decay,
 and the inelasticities in both cases are also 
quite different.} 
Of course, you can always give numbers that fit the experimentally observed moments 
in peripheral two-pion production; 
but so will other, in some cases very different values of $\delta$s and $\eta$s.
In the energy region   $1.4\leq s^{1/2}\leq 2$ GeV,  the  
phase shifts and inelasticities all stem from a single set of 
experiments and are likely to disagree with reality by much more than 
their nominal errors. 
In fact, this can be seen to occur for the S wave even at lower energy: 
 as soon as 
 the $\bar{K}K$ channel opens, 
the Cern-Munich phase shifts\ref{11} disagree violently with the Berkeley\ref{12} ones.
This  emphasizes the dangers of relying on a single
experiment for the
 phase shifts, as one has to do already for $s^{1/2}\geq1.2\,\gev$.

It is not difficult to see how   different phases may give similar results, for the 
 elastic cross section.
 For example, consider  the 
{\sl elastic} $\pi\pi$ cross section, in the P wave: in both cases 
(Cern-Munich and Particle Data Tables results)  
it is small. 
In the Cern-Munich one, because $\sin^2\delta_1$ is small; in the 
other because $\eta$ is small. 
Unfortunately, the {\sl imaginary} parts of the inelastic amplitudes are 
very different; contrary to the Cern-Munich 
results, in the PDG case it would be large, at least around the resonances, because 
of the contribution of the inelastic channels. 
The converse (i.e., overestimate of the total cross section)
 may, of course, also happen. 
In fact, the cases mentioned are just examples of an ambiguity 
(over and above that due to experimental errors) proved to exist quite generally in ref.~16, 
and which is likely to be large as soon as you have important inelastic channels open.

Now, CGL (following Pennington\ref{17}) take the Cern-Munich 
phase shifts, that probably contain large and unknown systematic errors, 
 and impose 
sum rules [e.g., the sum rules (B.6,7), (C.2) in ACGL], following from low energy crossing symmetry,
 to fix the Regge parameters at  energies $E>2\,\gev$. 
Not surprisingly,  CGL (and Pennington\fnote{This 
 should not be taken as a criticism of the work of Pennington; 
at that time the data, very poor, did indeed suggest possible deviations of 
Regge behaviour for $\pi\pi$. On the other hand, the fact that 
QCD also implies factorization was of course unknown. 
}) get irrealistic
Regge parameters (as realized by CGL themselves);
 for example, ACGL and CGL get a Pomeron with a width of the diffraction 
peak which is $s$-independent, and  
twice the   standard value (at low $s$), and 
a residue much smaller than what factorization implies.  
In fact, 
we will show in the Appendix  explicit 
calculations of two sum rules (in particular of  the sum rule (B.7), one of the  
crossing sum rules that Pennington and ACGL use)  which  
are perfectly satisfied by 
a standard Regge amplitude, with factorization for the rho 
and Pomeron trajectories, provided one uses 
Regge asymptotics from $s^{1/2}=1.42\,\gev$. 

According to CGL this 
deviation from conventional Reggeistics is not important 
because the influence of the 
high energy region ($s^{1/2}\geq1.42$ \gev) into their low energy 
 ($s^{1/2}\leq0.82$ \gev) phase shifts is very slight. 
However, and as we have shown in the present paper, 
inconsistencies show up as soon as one considers sum rules 
--like the Froissart--Gribov sum rules-- that are sensitive 
to the high energy behaviour of the amplitudes.

From the previous analysis it thus follows that CGL start, in the Roy equations,
 from a $V$ with incorrect Regge behaviour and dubious phase shifts 
above $1.42$ GeV. Let us call this $V({\rm Wrong\; R})$. 
CGL run this through the Roy equations (2.6) and find a solution, 
$\xi({\rm Wrong\; R})$. 
Now, this solution is not horrendous because experimental low energy data, 
chiral perturbation theory 
and crossing sum rules force you to 
have the errors in Regge parameters and cross sections compensating, to a 
certain extent,
 in what regards 
their low energy effects.  
Indeed, the independence on the low energy partial waves on the
high energy amplitudes used is approximately true, for the ACGL results, where
 the mismatch that occurs if using the correct Regge asymptotics stays
below the
$2\,\sigma$  level. 
However, for the {\sl CGL} results, use of 
 chiral perturbation theory 
(with neglect of higher order corrections) has the dual
 effect of highly correlating the various low energy parameters
 and excessively decreasing the errors. 
Thus, for example, the value for the quantity $a_{+0}$ that follows from the Froissart--Gribov 
representation, $10.94\pm0.13$ (in units of $10^{-4}M^{-5}_\pi$) is displaced 
$4\,\sigma$ from the value 
following directly from the parameters of CGL, $10.53\pm0.10$. 
Now,  $a_{+0}$ is directly related to the chiral constant $\bar{l}_2$,
 $a_{+0}=[\bar{l}_2-27/20]/720\pi^3f^4_\pi M_\pi$. Hence a variation of 
 $a_{+0}$ implies a corresponding variation of $\bar{l}_2$,  
or of higher chiral corrections, 
 that destabilizes all the quantities that
depend on it 
in a chiral perturbative analysis; in particular, the low energy S and P waves.
As we have shown in the present paper, inconsistencies show up 
in the CGL scattering amplitude (with standard Regge parameters) 
as soon as one considers sum rules that, like the Froissart--Gribov or Olsson 
ones, are sensitive to the high energy behaviour.
What the inconsistencies found in the previous section show is that the 
distortion is several times larger than the nominal CGL error bars.

\brochuresubsection{5.2. A tentative alternate solution}

\noindent
In support of the idea that 
 the effects discussed in the 
previous subsection are indeed the cause of the mismatches in the CGL 
$S$ matrix, we have calculated the Olsson sum rule and 
the quantities $a_l$, $b_1$  using now, for $s^{1/2}\leq0.82\,\gev$, the results of the fit, 
wave by wave, reported in ref.~6, \sect~7.6. 
For the wave S0 we take now the fit obtained imposing the value 
 $\delta_0^{(0)}(M_K^2)=43.3\pm2.3\degrees$, and with only three parameters;\fnote{This 
fit is actually a refinement of that of \equn{(7.6.2)} in ref~6; 
more details about this will be presented in a separate publication.} 
we  then have
$$\eqalign{
\cot\delta_0^{(0)}(s)=&\,\dfrac{s^{1/2}}{2k}\,\dfrac{M_{\pi}^2}{s-\tfrac{1}{2}M_{\pi}^2}\,
\dfrac{M^2_\sigma-s}{M^2_\sigma}\,
\left\{B_0+B_1\dfrac{\sqrt{s}-\sqrt{s_0-s}}{\sqrt{s}+\sqrt{s_0-s}}\right\};\cr
{B}_0=&\,21.04,\quad {B}_1=6.62,\quad
M_\sigma=782\pm24\,\mev;\cr
\dfrac{\chi^2}{\rm d.o.f.}=&\,\dfrac{15.7}{19-3};\quad 
a_0^{(0)}=(0.230\pm0.010);\quad\delta_0^{(0)}(M_K)=41.0\degrees\pm2.1\degrees.
\cr    }
\equn{(5.4a)}$$
The errors of the $B_i$ are strongly correlated; uncorrelated errors are obtained if 
replacing the $B_i$ by the 
parameters $x,\,y$ with
$$B_0=y-x;\quad B_1=6.62-2.59 x.
\equn{(5.4b)}$$
Then,
$$y=20.04\pm0.75,\quad x=0\pm 2.4.
\equn{(5.4c)}$$
The solution is shown, compared to the CGL phase, in Fig.~5.1.
We then integrate with (5.4) up to $E=0.82\;\gev$ and with (3.2) from $0.82s$ to $\bar{K}K$ 
threshold.
For S2, P we take the same fits as before, specifically, eqs. (3.3), (3.5).

\topinsert{
\setbox0=\vbox{\hsize13.truecm{\epsfxsize 10.8truecm\epsfbox{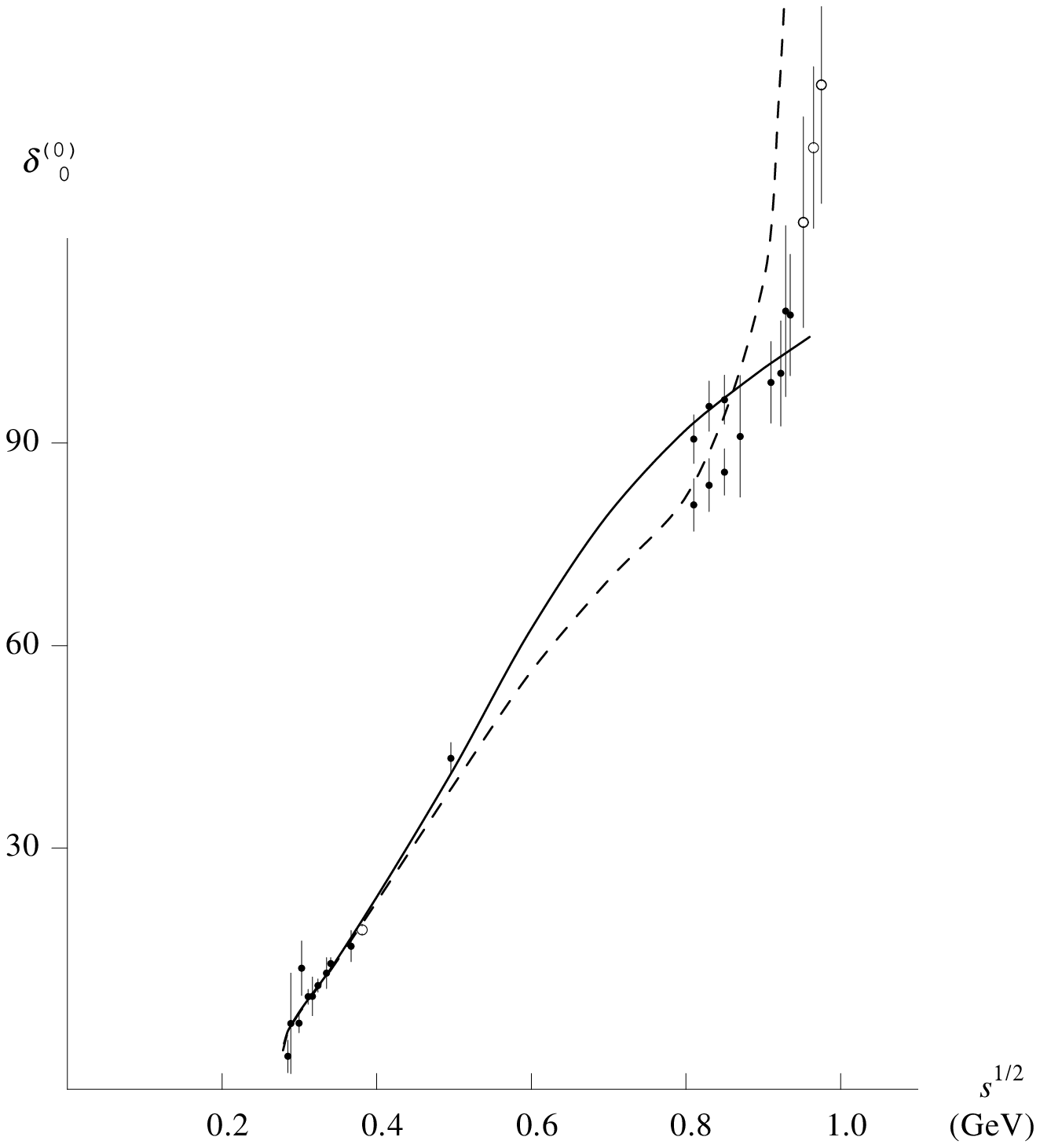}}} 
\setbox6=\vbox{\hsize 14truecm\captiontype\figurasc{Figure 5.1. }{
The  
$I=0$, $S$-wave phase shifts corresponding to (5.4) (continuous line) and 
 Colangelo, Gasser and Leutwyler, ref.~2 (dashed line). 
Some experimental points are also shown.}\hb} 
\centerline{\tightboxit{\box0}}
\bigskip
\centerline{\box6}
\medskip
}\endinsert

We find the following results, in units of $M_\pi$:
\smallskip
$$\matrix{
{\rm Olsson}&{\rm direct}&{\rm dispersive}\cr
&0.691\pm0.042&0.659\pm0.020\cr
}
\equn{(5.5)}$$
[here ``direct" means that we take the values following from the fits in  Eqs.~(3.3), (5.4)]. 
Moreover, and also in units of $M_\pi$,
\smallskip
$$\matrix{
a_1&{\rm direct,\;(TY,\;St.)}&{\rm direct,(TY,\;St. + Sys.)}&{\rm Froissart-Gribov}\cr
&(40.6\pm1.4)\times10^{-3}&(38.6\pm1.2)\times10^{-3}&(37.9\pm1.4)\times10^{-3}\cr
}
\equn{(5.6)}$$
and
\smallskip
$$\matrix{
b_1&{\rm direct,\;(TY,\;St.)}&{\rm direct,\;(TY,\;St. + Sys.)}&{\rm Froissart-Gribov}\cr
&(4.18\pm0.43)\times10^{-3}&(4.47\pm0.29)\times10^{-3}&(5.69\pm0.96)\times10^{-3}.\cr
}
\equn{(5.7)}$$
\smallskip\noindent
The tag ``direct" now refers to the values of ref.~10, with only statistical errors (St.) or 
including also systematic errors (St. + Sys.). 
Thus, we find agreement at the $1\,\sigma$ level in all three cases; 
for $a_1$, $b_1$, with the TY~(St.+Sys.) solution.
With the same parameters we find, for the D waves, and with 
the help of the Froissart--Gribov representation, the values
$$a_{0+}=(10.60\pm0.17)\times10^{-4}\;M_{\pi}^{-5},\quad
 a_{00}=(14.99\pm0.68)\times10^{-4}\;M_{\pi}^{-5}
\equn{(5.8a)}$$
and
$$b_{0+}= (-0.170\pm0.083)\times10^{-4}\,M_{\pi}^{-7},\quad 
b_{00}= (-6.91\pm0.47)\times10^{-4}\,M_{\pi}^{-7}.
\equn{(5.8b)}$$
This is compatible with what we found for the $a_2^{(I)}$
 by direct fit to the experimental 
data in \sect~3.3 within the rather large errors of these last values. 

The large error, and the separation in the central values in 
the Olsson sum rule, Eq.~(5.5), is due to the 
fact that the data do not fix with sufficient accuracy the $a_0^{(2)}$ 
scattering length, which 
provides most of the error in the ``direct" number. 
In fact, as is known, one can use the Olsson sum rule to refine the parameters of the S2 wave; 
if we do so, fixing all other parameters to their 
central values (within errors) and include the Olsson sum rule 
in the fit to the S2 wave we find
\smallskip
$$\eqalign{
\cot\delta_0^{(2)}(s)=&\,\dfrac{s^{1/2}}{2k}\,\dfrac{M_{\pi}^2}{s-2z_2^2}\,
\left\{B_0+B_1\dfrac{\sqrt{s}-\sqrt{s_0-s}}{\sqrt{s}+\sqrt{s_0-s}}\right\};\cr
s_0^{1/2}=1.45\;\gev;&\quad\chi^2/{\rm d.o.f.}=17.2/(19-2).\cr
 B_0=&\,-118\pm2.5,\quad B_1=-105\pm2.5,\quad z_2=139.57\;\mev\;\hbox{[fixed]}.\cr
}
\equn{(5.9)}$$
Then one has $a_0^{(2)}=-0.0428\pm0.0022$ and (5.5) becomes
$$\matrix{
{\rm Olsson}&{\rm direct}&{\rm dispersive}\cr
&0.671\pm0.023&0.663\pm0.018.\cr
}
\equn{(5.10)}$$
The rest of the relations (5.6-8)  improve slightly, and 
the D wave scattering lengths also change a little:
$$\eqalign{a_1=&\,38.0\pm1.2\,\times10^{-3}\,M_{\pi}^{-3},\quad
b_1=5.64\pm0.96\,\times10^{-3}\,M_{\pi}^{-5};\cr 
a_{0+}=&\,(10.51\pm0.15)\times10^{-4}\;M_{\pi}^{-5},\quad
 a_{00}=(14.89\pm0.65)\times10^{-4}\;M_{\pi}^{-5}.\cr
}
\equn{(5.11)}$$
It should be noted that the error here for $a_{0+}$
 is at the edge of the region of credibility, as  
indeed it is of the order of magnitude of electromagnetic corrections which the analysis does 
not take into account. 
This value of $a_{0+}$ implies, at one loop level, a very precise value for the 
chiral perturbation theory parameter\ref{19} $\bar{l}_2$ of
$$\bar{l}_2=5.97\pm0.07$$

Of course the agreement in (5.5,\/6,\/7,\/10) is not enough to guarantee
 that the new solution is consistent; 
to prove that, 
one would have  to check the whole set of dispersion relations and crossing constraints, 
something that will be the subject of a separate paper. 
But it clearly suggests that the CGL solution fails to pass the tests because 
it is distorted. 
This can   also be inferred by comparing the CGL solution for the 
S2 wave with (3.3) as in  Fig.~(5.2), where we show the CGL and  (3.3) together. 
While both fit the data below 0.82 \gev\ [expression (3.3) gives actually a slightly 
better fit even there], the distortion
 of the CGL solution above that energy is suggestive. 
A similar pattern is found in Figs.~3.1, 5.1. 
This very much suggests that the CGL fit is a forced fit, biased 
by a reflection of a faulty high energy scattering amplitude.

\topinsert{
\setbox0=\vbox{\hsize15.4truecm{\epsfxsize 14truecm\epsfbox{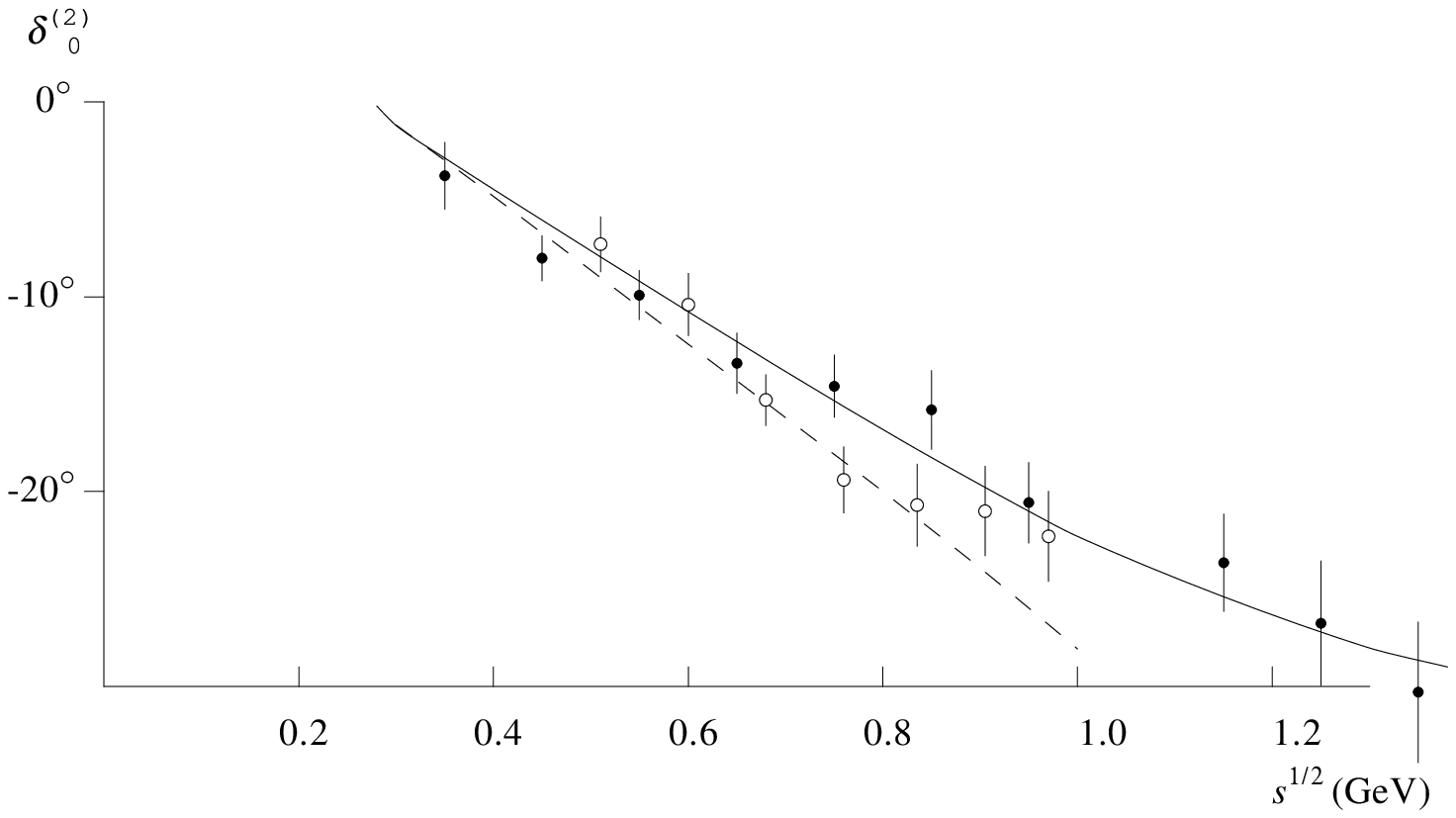}}} 
\setbox6=\vbox{\hsize 11truecm\captiontype\figurasc{Figure 5.2. }{
The  
$I=2$, $S$-wave phase shifts corresponding to (3.3) (continuous line) and 
 Colangelo, Gasser and Leutwyler, ref.~2 (dashed line). 
Also shown are the data points of losty et al. (open circles) and 
 from solution~A of Hoogland et al. (black dots), refs.~13.}\hb} 
\centerline{\tightboxit{\box0}}
\bigskip
\centerline{\box6}
\medskip
}\endinsert

\booksection{6. Summary and conclusions}

\noindent
We have checked a number of  tests of the low energy
($s^{1/2}\leq0.82$ \gev) S0, S2 and P wave phase shifts 
given in ref.~2 by Colangelo, Gasser and Leutwyler, based on two loop 
chiral perturbation theory plus the Roy equations with a 
certain high energy ($s^{1/2}\geq1.42\,\gev$) input. 
We have shown that, if we used the values for this high energy 
piece that follow from Regge theory, then the Olsson sum rule and the combinations of scattering 
lengths $a_{0+}=\tfrac{2}{3}[a_2^{(0)}-a_2^{(2)}]$, $a_{00}=\tfrac{2}{3}[a_2^{(0)}+2a_2^{(2)}]$
 show mismatch by 
as much as  $4\,\sim5\,\sigma$.
We have discussed in detail why we think that 
the discrepancy is inherent to the {\sl low energy} ($s^{1/2}\leq0.82$ \gev) CGL phases. 
Thus, in \subsect~4.4.1 we have shown that even rather
 drastic alterations of the middle energy region, $0.82\leq s^{1/2}\leq1.42$ do 
not alter the inconsistencies.

With respect to the higher energy region ($s^{1/2}\geq1.42$ \gev),  
the situation is such that, if one tries to modify the Regge piece 
to fit the Olsson sum rule (say) then not only the alteration (40 to 100\%) is much more 
than what one can reasonably expect, but  
the lack of consistency is shifted to $a_1,\,b_1$.
A similar phenomenon --in fact, even more pronounced-- occurs with $a_{0+}$ 
and $a_{00}$. 
This we discussed in detail in \subsect~4.4.2, 
where it is clear that the mismatch is due to the low energy 
CGL input.
Moreover, the value of the quantity 
$b_1$ remains displaced by $4\,\sigma$ from what one gets 
from a fit to the pion form factor.

It should be borne in mind that we are talking here about disagreements at 
the level of a few percent; so, if one is prepared to shift the 
central values of CGL by up to $2\,\sigma$, and double their errors,
 the inconsistencies disappear. This is what happens, for example in the analysis of ACGL 
where the errors are from 3 to 10 times larger than those in CGL.
Nevertheless, at the level of precision claimed by CGL, the errors are real. 
We have argued that they are probably due to an irrealistic high energy  ($s^{1/2}\geq1.42$ \gev)
input, which distorts the low energy phase shifts. 
In support of this we have shown that a direct fit to data, 
including fully analyticity constraints, for the 
P, S0, S2 waves (in the case of the last, 
requiring also consistency of the Olsson sum rule to decrease its errors)
 plus a high energy input 
given by orthodox Regge theory, produces a different set
 of  compatible low energy phase shifts and high energy scattering amplitude. 
This set is formed by the phase shifts given in Eqs.~(3.5), (5.4) 
and (5.9). 
This set is in fact similar to that of CGL, but is slightly displaced
 and its errors are slightly larger;
so, for example, the quantity $\delta_0^{(0)}(M^2_K)-\delta_0^{(2)}(M^2_K)$, 
important for kaon decays, changes according to
$$\delta_0^{(0)}(M^2_K)-\delta_0^{(2)}(M^2_K)=47.7\pm1.5\degrees\;\hbox{[CGL]}\;
\to\;48.4\pm2.1\degrees\;
\hbox{[Our solution]}.$$

A fact that may be mentioned here is that Descotes et a.\ref{2} 
have, in a recent article, found a solution whose central 
values differ from that of CGL by almost 2(CGL) standard deviations 
and in fact point in the direction of our tentative alternate solution here. 
Thus, they have, in units of $M_\pi$, 
$$a_0^{(0)}=0.228\pm0.012,\;a_0^{(2)}=-0.0382\pm0.0038\;[\hbox{Descotes et al.}]$$ 
Their errors are also more like what we have in our alternate solution. 
Note, however, that whether or not the alternate solutions turn out to  
be consistent has nothing to do with the consistency of the CGL solution: 
this last fails independently of the failure or success of the 
novel one(s).

Analyticity determines the real part of 
the scattering amplitude in terms of its imaginary part.  
However, to get the real part you need to know the imaginary 
part up to infinity. Now, if the  imaginary part is wrong 
at high energy and yet the dispersion relation 
(or Roy equations) are satisfied,  
it necessarily follows that one must have 
made a compensating error in the low energy 
imaginary part. 
In other words: you 
have fallen into a spurious solution.
The fact that the solution is spurious should be manifest as soon as 
one devises a test that gives a {\sl different} weight to high and low 
energy pieces: this is exactly what we do  in our paper, 
for the 
CGL solution, with 
the help of the Froissart-Gribov representations. 

\booksection{Appendix}

\noindent
In this Appendix we briefly discuss (and prove the failure of) the reason for the unorthodox 
Reggeistics chosen in ACGL, CGL,  
in as much as it has a bearing on our subject matter here. 
These authors, following Pennington,\ref{17} set up crossing sum rules [Eqs.~(B.7), (C.2) in ACGL], 
which relate high and low energy, and conclude that they are satisfied only if, 
in particular, the Pomeron residue is about \ffrac{1}{3} of the 
value implied by factorization and the rho residue is about a 40\% larger.
 
Contrarily to the conclusion of ACGL, however, we will show 
by explicit calculation of two typical sum rules that, if 
one assumes orthodox Regge behaviour from $s^{1/2}\geq1.42\,\gev$, 
the low energy phase shifts are perfectly compatible with 
the value of the Regge residues implied by factorization. 
This will cinch the proof that, as discussed in \subsect~5.1, the 
 Reggeistics of ACGL are very likely due to compensation of the unrealistic
  phase shifts 
used between $1.42\leq s^{1/2}\leq2\,\gev$.

Specifically, we will consider a sum rule dominated at high energy 
by the Pomeron, viz., the sum rule (B.7) in ACGL; since 
it is independent of the S and P waves, it constitutes a new, stringent test 
of the Regge structure. We will also consider another sum rule, 
which is dominated by the rho trajectory. 

The first sum rule may be written as
$$J\equiv\int_{4M^2_\pi}^\infty\dd s\,\Bigg\{
\dfrac{4\imag F'^{(0)}(s,0)-10\imag F'^{(2)}(s,0)}{s^2(s-4M^2_\pi)^2}
-6(3s-4M^2_\pi)\,\dfrac{\imag F'^{(1)}(s,0)-\imag F^{(1)}(s,0)}{s^2(s-4M^2_\pi)^3}
\Bigg\}=0.
\equn{(A.1)}
$$
Here $F'^{(I)}(s,t)=\partial F^{(I)}(s,t)/\partial\cos\theta$, 
and the   index $I$ refers to isospin in the $s$ channel.

We will separate $J$ into a low energy and a high energy piece:
$$J=J_{\rm l.e.}+J_{\rm h.e.};\quad
J_{\rm l.e.}=\int_{4M^2_\pi}^{s_h}\dd s\,\dots, \quad
J_{\rm h.e.}=\int_{s_h}^\infty\dd s\,\dots\,. 
$$ 
The low energy piece, $J_{\rm l.e.}$, only contains contributions of waves D 
and higher. Since these waves are only known with (relatively) 
large errors,\fnote{These errors 
are particularly large, and uncertain,  
above 1.3 \gev, where inelasticity begins to be important. 
For example, the error on the D0 wave contribution to 
$J_{\rm l.e.}$ due to a 50\% change in the 
inelasticity of the $f_2(1270)$ resonance is 
as large as the nominal error 
due to only the errors in the 
parameters in (3.9).} it is 
(generally speaking) very dangerous to 
draw conclusions about the high energy integral,  $J_{\rm h.e.}$, from the 
experimental value of the low energy piece, $J_{\rm l.e.}$. 
Nevertheless, we will show that, if we choose $s_h=1.42^2\,\gev^2$, 
then we find perfect consistency, within errors. 
In this calculation we will first neglect the contributions of 
exchange of $I=2$ and of the background to rho exchange, 
both of dubious status and substantially smaller than the  Pomeron and rho exchange pieces, 
but we keep the  $P'$. 
Using the parametrizations of \sect~3.3 for the D, F waves  
we find, in units of $M_{\pi}^{-6}$,
$$J_{\rm l.e.}({\rm D\; waves})=1.222\times10^{-4},\quad
J_{\rm l.e.}({\rm F\; wave})=-0.076\times10^{-4}
$$
so that, including the errors,
$$J_{\rm l.e.}=(1.15\pm0.05)\times10^{-4}.
\equn{(A.2)}$$
For the high energy piece, expanding in amplitudes with definite isospin in the 
$t$ channel, and with the numbers in \sect~3.4 for the 
Pomeron and rho contributions, we get  
$$J_{\rm h.e.}({\rm Pomeron})=-1.093\times10^{-4},
\quad J_{\rm h.e.}(\rho)=0.034\times10^{-4}, 
$$
i.e., including errors, 
$$J_{\rm h.e.}=(-1.06\pm0.17)\times10^{-4}. 
\equn{(A.3)}$$
Thus, we have cancellation between (A.2) and (A.3),  
within errors:
 there is no reason  to justify departure off the expected Regge behaviour.

We next comment a little on the $P'$ and on the inclusion of 
the $I_t=2$ contribution. Because the high energy part of the sum rule (A.1) 
is mostly given by the $t$ derivative of the even isospin amplitudes, 
a more precise evaluation than the one carried here would require 
that we replace the $P'$ contribution of (3.17a) by a more accurate formula. 
Unfortunately, the characteristics of this Regge pole are poorly known; see 
ref.~14. If we take for the 
 the $P'$ trajectory  a formula like that of the 
$\rho$, then 
 (A.3) is replaced by
$$J_{\rm h.e.}(\hbox{With corrected}\,P')=(-1.2\pm0.2)\times10^{-4}. $$
Including also the $I_t=2$ contribution, as given in (3.18), 
we would find 
$$J_{\rm h.e.}(\hbox{With corrected}\,P',\,\hbox{and including}\,I_t=2) =
(-0.5\pm0.3)\times10^{-4}.
\equn{(A.4)}$$
This  still cancels the low energy piece, (A.2) at the $2\,\sigma$ 
level. This  discrepancy cannot be taken seriously, because of the 
uncertainties in the $P'$ trajectory and because  
the $t$  slope in formula (3.18) is little more than guesswork.

The second sum rule is obtained by profiting from the threshold behaviour to 
write an unsubtracted forward dispersion relation for the 
quantity $F^{(I_s=1)}(s,0)/(s-4M^2_\pi)$ 
This gives the relation
$$\dfrac{6 M_\pi}{\pi}a_1=\dfrac{1}{\pi}\int_{M^2_\pi}^\infty\dd s\,
\dfrac{\imag F^{(I_s=1)}(s,0)}{(s-4M^2_\pi)^2}+\dfrac{1}{\pi}\sum_IC^{(su)}_{1I}
\int_{M^2_\pi}^\infty\dd s\,
\dfrac{\imag F^{(I)}(s,0)}{s^2},
\equn{(A.5)}$$
 which is known at times as the (second) Olsson sum rule.
The index $I$ refers to isospin in the $s$ channel and $C^{(su)}_{1I}$ are the $s-u$ 
crossing matrix elements.
Canceling $a_1$ with the Froissart--Gribov expression for this quantity 
and substituting the $C^{(su)}_{1I}$ we find the result
$$I\equiv \int_{M^2_\pi}^\infty\dd s\,
\dfrac{\imag F^{(I_t=1)}(s,0)-\imag F^{(I_t=1)}(s,0)}{s^2}-
 \int_{M^2_\pi}^\infty\dd s\,\dfrac{8M^2_\pi[s-2M^2_\pi]}{s^2(s-4M^2_\pi)^2}
\imag F^{(I_s=1)}(s,0)\equiv I_1+I_2=0.
\equn{(A.6)}$$

The contributions of the S waves cancel in (A.6), so only the P, D and F waves 
contribute (as usual, we neglect waves G and higher). 
At high energy, $I_2$ contributes little since the corresponding integral converges rapidly: 
most of the high energy contribution comes from the first term, dominated by rho exchange.
We will use units so that $M_\pi=1$ and obtain the following results:
$$\eqalign{
I(\hbox{low energy, P wave})=(-2.80\pm0.31)\times10^{-2},\cr
I(\hbox{low energy, D0 + D2 waves})=(0.56\pm0.03)\times10^{-2},\cr
I(\hbox{low energy, F wave})=(0.01\pm0.00)\times10^{-2},\cr
I(\hbox{high energy}, \rho)=(2.41\pm0.37))\times10^{-2},\cr
I(\hbox{high energy}, I=0)=-(0.17\pm0.02)\times10^{-2}\cr
I(\hbox{high energy}, I=2)=-(0.02\pm0.01)\times10^{-2}.\cr
}
\equn{(A.7)}$$
By ``low energy" we understand the contributions from energies below 
$1.42\,\gev$, where we use phase shifts and inelasticities to calculate the scattering 
amplitudes, and ``high energy" is 
above 1.42 \gev, where a Regge description is employed.
The final result for the sum rule is
$$I=(0.016\pm0.37)\times 10^{-2},$$
i.e., complete cancellation of low and high energy contributions.

The remarkable fulfillment of these sum rules show the incorrectness
 of the assertions found  in ACGL, CGL: both for Pomeron and rho,   
standard Regge behaviour for $\pi\pi$ scattering is perfectly consistent 
with crossing symmetry provided one imposes it systematically for energies above 1.42 \gev.

\booksection{Note added in proof}

\noindent
After this article was sent to the publisher, a preprint has appeared [I.~Caprini, G.~Colangelo, 
J.~Gasser and H.~Leutwyler, hep-ph/0306122] 
in which some of the conclusions (but not the calculations) of our work are contested.
We do not think it necessary to alter our paper on account of the work of Caprini et al.; 
we plan to present a discussion of it in a separate article.

\vfill\eject
\brochuresection{Acknowledgments}

\noindent
One of us (FJY) is grateful to 
G.~Colangelo, J.~Gasser and H.~Leutwyler for most interesting discussions that 
have triggered his interest in this subject. Both of us thank again 
Prof.~Leutwyler for a very useful critical reading of a preliminary draft 
which, in particular, allowed us to correct a few inconsistencies. 
J.R.P. acknowledges 
support from the Spanish CICYT projects
PB98-0782 and BFM2000 1326, as well as a
Marie Curie fellowship MCFI-2001-01155; FJY is grateful to INTAS for partial financial support.

\brochuresection{References}
\item{1 }{Ananthanarayan, B., et al., {\sl Phys. Rep.}, {\bf 353}, 207,  (2001).}
\item{2 }{Colangelo, G., Gasser, J.,  and Leutwyler, H.,
 {\sl Nucl. Phys.} {\bf B603},  125, (2001).}
\item{3 }{Roy, S. M., {\sl Phys. Letters}, {\bf 36B}, 353,  (1971).}
\item{4 }{Palou, F. P., and Yndur\'ain, F. J., {\sl Nuovo Cimento}, {\bf 19A}, 245, 
 (1974).}
\item{5 }{Palou, F. P., S\'anchez-G\'omez, J. L., and Yndur\'ain, F. J., 
{\sl Z. Phys.}, {\bf A274}, 161, (1975).}
\item{6 }{Yndur\'ain, F. J. ,  ``Low energy pion interactions", FTUAM02-28 (hep-ph/0212282).}
\item{7 }{Martin, B. R., Morgan, D., and and Shaw, G., {\sl 
Pion-Pion Interactions in Particle Physics}, Academic Press, New York, (1976).}
\item{8 }{Atkinson, D., {\sl Nucl. Phys.} {\bf B7}, 375  (1968) and  
 {\sl Nucl. Phys.}, {\bf B23}, 397 (1970).}
\item{9 }{For the more recent determination, 
see Aloisio, A., et al., {\sl Phys. Letters}, {\bf B538}, 21,  (2002); the older one is from 
Pascual, P., and Yndur\'ain, F. J., {\sl Nucl. Phys.} {\bf B83}, 362, (1974).} 
\item{10 }{de Troc\'oniz, J. F., and Yndur\'ain, F. J., {\sl Phys. Rev.},  {\bf D65}, 093001,
 (2002).}
\item{11 }{Hyams, B., et al., {\sl Nucl. Phys.} {\bf B64}, 134, (1973); 
Grayer, G., et al.,  {\sl Nucl. Phys.}  {\bf B75}, 189, (1974).. See also the analysis of the 
same experimental data in
Estabrooks, P., and Martin, A. D., {\sl Nucl. Physics}, {\bf B79}, 301,  (1974).}
\item{12 }{Protopopescu, S. D., et al., {\sl Phys Rev.} {\bf D7}, 1279, (1973).}
\item{13 }{Losty, M.~J., et al.  {\sl Nucl. Phys.}, {\bf B69}, 185, (1974); 
Hoogland, W., et al.  {\sl Nucl. Phys.}, {\bf B126}, 109, (1977).}
\item{14 }{Gell-Mann, M.  {\sl Phys. Rev. Letters}, {\bf 8}, 263, (1962); 
Gribov, V. N., and Pomeranchuk, I. Ya.  {\sl Phys. Rev. Letters}, {\bf 8}, 343,  (1962). 
For more references in general 
Regge theory, see Barger, V. D., and Cline, D. B., {\sl Phenomenological Theories of High Energy 
Scattering}, Benjamin, New~York,  (1969); for references to the QCD 
analysis, Yndur\'ain, F. J.,  {\sl The Theory of Quark and 
Gluon Interactions}, Springer, Berlin, 1999.}
\item{15 }{Rarita, W., et al., {\sl Phys. Rev.} {\bf 165}, 1615, (1968).}
\item{16 }{Particle Data Tables: D. E. Groom et al., {\sl Eur. Phys. J.} {\bf C15}, 1, (2000).}
\item{17 }{Atkinson, D., Mahoux, G., and Yndur\'ain, F. J., {\sl Nucl. Phys.} {\bf B54}, 263, 
 (1973); 
{\bf B98}, 521 (1975).}
\item{18 }{Pennington, M. R., {\sl Ann. Phys.} (N.Y.), {\bf 92}, 164, (1975).}
\item{19 }{Gasser, J., and Leutwyler, H., {\sl Ann. Phys.} (N.Y.), {\bf 158}, 142,  (1984).}
\item{20 }{Descotes, S., Fuchs, N. H.,  Girlanda, L., and   Stern, J., {Eur. Phys. J. C}, 
{\bf 24}, 469, (2002).}

\bye